\documentclass[sigconf,9pt]{acmart}

\copyrightyear{2018}
\acmYear{2018}
\setcopyright{acmcopyright}
\acmConference[ICN '18]{ICN '18: 5th ACM Conference on Information-Centric Networking}{September 21--23, 2018}{Boston, MA, USA}
\acmBooktitle{ICN '18: 5th ACM Conference on Information-Centric Networking (ICN '18), September 21--23, 2018, Boston, MA, USA}
\acmPrice{15.00}
\acmDOI{10.1145/3267955.3267967}
\acmISBN{978-1-4503-5959-7/18/09}

\newcommand{\eg}{\textit{e.g.,}~}
\newcommand{\ie}{\textit{i.e.,}~}

\usepackage{xspace}

\newcommand{\one}{({\em i})\xspace}
\newcommand{\two}{({\em ii})\xspace}
\newcommand{\three}{({\em iii})\xspace}

\usepackage{geometry}

\usepackage[utf8]{inputenc}
\usepackage[T1]{fontenc}
\usepackage{lmodern}

\usepackage{balance}

\usepackage{color}
\usepackage{subfigure}
\usepackage{booktabs}
\usepackage{makecell}
\usepackage{multirow}
\usepackage{enumitem}
\usepackage[english]{babel}

\usepackage[absolute,showboxes]{textpos}

\usepackage{tikz}
\usetikzlibrary{calc,colorbrewer,decorations,decorations.pathmorphing,decorations.pathreplacing,fit,positioning,shapes,shapes.misc}

\usepackage{pifont}
\newcommand{\cmark}{\ding{51}}%
\newcommand{\xmark}{\ding{56}}%

\graphicspath{{figs/}}

\makeatletter
\renewcommand{\paragraph}[1]{\vspace*{0.03in}\noindent{\bf #1.}\hspace{0.25ex \@plus1ex \@minus.2ex}}
\makeatother

\begin{document}

\title{The Need for a Name to MAC Address Mapping: Quantifying Resource Consumptions of an Uninformed Data Link Layer}
\title[NDN  in the IoT]{NDN, CoAP, or MQTT --- who Fuels  the Better IoT? \\ A Comparative Measurement Study}
\title[NDN, CoAP, and MQTT in the IoT]{NDN, CoAP, and MQTT: A Comparative \\ Measurement Study in the IoT}

\author{Cenk G{\"u}ndo\u{g}an}
\affiliation{%
  \institution{HAW Hamburg}
}
 \email{cenk.guendogan@haw-hamburg.de}

\author{Peter Kietzmann}
\affiliation{%
  \institution{HAW Hamburg}
}
 \email{peter.kietzmann@haw-hamburg.de}

\author{Martine Lenders}
\affiliation{%
  \institution{Freie Universit\"at Berlin}
}
 \email{m.lenders@fu-berlin.de}

 \author{Hauke Petersen}
\affiliation{%
  \institution{Freie Universit\"at Berlin}
}
 \email{hauke.petersen@fu-berlin.de}
 
\author{Thomas C. Schmidt}
\affiliation{%
  \institution{HAW Hamburg}
}
 \email{t.schmidt@haw-hamburg.de}

\author{Matthias W{\"a}hlisch}
\affiliation{%
  \institution{Freie Universit\"at Berlin}
}
\email{m.waehlisch@fu-berlin.de}

\renewcommand{\shortauthors}{C. G{\"u}ndo\u{g}an et al.}

\begin{abstract}

	This paper takes a comprehensive view on the protocol stacks that are under debate for a future Internet of Things (IoT).
	It addresses the holistic question of which solution is beneficial for common IoT use cases. We deploy NDN and the two popular IP-based application protocols, CoAP and MQTT, in its different variants on a large-scale IoT testbed in single- and multi-hop scenarios. We analyze the use cases of scheduled periodic and unscheduled traffic under varying loads. Our findings indicate that (a) NDN admits the most resource-friendly deployment on nodes, and (b) shows superior robustness and resilience in multi-hop scenarios, while (c) the IP protocols operate at less overhead and higher speed in single-hop deployments. 
Most strikingly we find that NDN-based protocols are in significantly better flow balance than the UDP-based IP protocols and require fewer corrective actions.
\end{abstract}



\maketitle

\setlength{\TPHorizModule}{\paperwidth}
\setlength{\TPVertModule}{\paperheight}
\TPMargin{5pt}
\begin{textblock}{0.8}(0.1,0.02)
     \noindent
     \footnotesize
     If you cite this paper, please use the ICN reference:
     C. G{\"u}ndo\u{g}an, P. Kietzmann, M. Lenders, H. Petersen, T.~C.
Schmidt, M. W\"ahlisch. NDN, CoAP, and MQTT: A Comparative
Measurement Study in the IoT. In \emph{Proc. of ACM ICN}, ACM, 2018.
\end{textblock}

\section{Introduction}
\label{sec:intro}

The Internet of Things (IoT) is evolving and an increasing number of  controllers in the field is augmented with network interfaces that speak IP.  Current deployments often are part of larger systems (e.g., a  heating)  or  attached to infrastructure (e.g., smart city lighting). Such devices connect to power, use common broadband links, and adopt the old MQTT protocol~\cite{mqtt311} for publishing IoT data to a remote cloud. The prevalent  use case forecasted for the IoT, though, consists of billions of constrained sensors and actuators mainly  not cabled to power, but  connected via low power lossy wireless links. The key target of the IoT will be data, of which a total of 1.6 Zettabytes is expected in 2020 \cite{ms-eai-15}. The mass constituents of the IoT will be tiny, cheap {\em things} that are severely challenged by the current way of connecting to the Internet.

This new class of connected devices cannot be integrated into today's Internet infrastructure without technologies that bridge the scale. The IETF has designed a suite of  protocols for successfully serving the needs of a constrained IoT.
 IPv6 adaptation layers such as 6LoWPAN \cite{RFC-4944} enable a deployment on constraint links (e.g., IEEE 802.15.4), which RPL routing  arranges in a multi-hop  topology \cite{RFC-6550}. The Constrained Application Protocol (CoAP) \cite{RFC-7252} offers a lightweight alternative to HTTP while running over UDP, or DTLS \cite{RFC-6347} for session security. This set of solutions extends the host-centric end-to-end paradigm of the Internet to the embedded world and puts IPv6 in place for loosely joining the {\em things}.

 However,  doubts arose whether host-to-host sessions are the appropriate approach in these disruption-prone environments of (wireless) things, and the data-centric nature at the Internet edge called for  rethinking the current IoT architecture \cite{szsmb-avdir-17}. ICN networks \cite{adiko-sind-12} have been identified as promising candidates to replace the rather complex IETF network stack in a future IoT. Name-based routing and in-network caching as contributed by Named Data Networking (NDN) \cite{jstp-nnc-09,zabjc-ndn-14} bear the potential to increase robustness of application scenarios in regimes of low reliability and reduced infrastructure (e.g., without DNS). Following initial concepts  \cite{olg-ccnte-10} and early experimental work \cite{bmhsw-icnie-14}, the adaptation, analysis, and deployment of NDN for the IoT became an active  research area that advocated  the IoT as a candidate of early NDN adoption. Still open problems persist, namely naming, routing, forwarding \cite{wsv-bdpts-13}, and data push  \cite{RFC-7927} as Shang et al. \cite{sblwy-ndnti-16} recently reminded. 

While the IETF and the ICN community tweak their protocols and companies deploy MQTT-to-cloud uplinks, the quest for the best solution remains open. Rather little is known about the differences and commonalities when deploying the varying approaches in the wild. This surprisingly unsatisfying state of the art motivates us to implement, deploy, and thoroughly analyze the different protocols in typical use cases and scenarios for the constrained Internet of Things.  

The main contribution of this paper is a  thorough comparative analysis of the three protocol families NDN, CoAP, and MQTT\footnote{We use the UDP-adapted version MQTT-SN, since TCP is inappropriate for the constrained IoT.} covering its main variants. We implemented characteristic IoT use cases for ten variants of these protocols,  deployed them in single- and multi-hop scenarios on a large IoT testbed, and ran competitive performance contests under fully equivalent conditions. Our  analysis showed common behavior for pull versus push solutions in single-hop experiments, but revealed significant differences in the challenging multi-hop domain. Flow performance, reliability and stability attained superior results for hop-wise replicated NDN protocols, while end-to-end approaches showed severe shortcomings at iterated link stress.

The remainder of this paper is structured as follows. The following Section \ref{sec:background} summarizes the key protocol properties and introduces the use cases. 
Section \ref{sec:setup} explains our implementations and experimental setup. 
We present  measurements and analyze the results in Section~\ref{sec:eval}, 
In Section~\ref{sec:related_work}, we revisit the related work
and conclude in Section~\ref{sec:c+o}.

\section{Background and Use Cases}
\label{sec:background}

\subsection{CoAP}

CoAP, the Constrained Application Protocol~\cite{RFC-7252}, was designed to support REST services in machine to machine communication.
Basically, it aims for replacing HTTP on constrained nodes.
In contrast to HTTP, CoAP is able to run on top of UDP and introduces a lean transactional messaging layer to compensate for the connectionless transport.
CoAP provides a more compact header structure than HTTP.

Three communication primitives are currently supported by this extensible protocol: 
\one pull, \two push, and \three observe.
Pull implements the common request response communication pattern.
However, as IoT scenarios also include the pro-active communication of unscheduled state changes, CoAP was extended to support pushing new events to its peers.
Still, this does not allow for publish-subscribe scenarios when producer and consumer are decoupled in time and data is not yet available at the request.
The support for delayed data delivery in publish-subscribe was specified in CoAP observe~\cite{RFC-7641}.
Here, clients can signal interest in observing data, which basically means that a CoAP server delivers data as soon as available and maintains state until clients explicitly unsubscribe.

CoAP must be considered as the IETF standard to implement application layer data transfer in the future Internet of Things. Currently, several implementations exist, as well as early adoption in a few selected products and deployments.

\subsection{MQTT}

MQTT~\cite{mqtt311}, the Message Queue Telemetry Transport, was designed as a publish-subscribe messaging protocol between clients and brokers.
Clients can publish content, subscribe to content, or both.
Servers (commonly called \emph{broker}) distribute messages between publishing and subscribing clients.
It is worth noting that the protocol is symmetric: Clients as well as brokers can be sender and receiver when MQTT delivers application messages.

MQTT is considered  a lightweight protocol for two reasons.
First, it provides a lean header structure, which reduces packet parsing and makes it suitable for IoT devices with low energy resources.
Second, it is easy to implement.
In its simplest form, MQTT offloads reliability support completely onto TCP.

To provide flexible Quality of Service on top of the underlying transport, MQTT defines three QoS levels, which reflect the agreement regarding message transfer between broker and consumer -- both can be sender and receiver.
\emph{QoS~0} implements unacknowledged data transfer.
An MQTT receiver gets a message at most once, depending on the capabilities of the underlying network, as there is no retransmission on the application layer.
\emph{QoS~1} guarantees that a message is delivered at least once.
This requires that a message is stored at the sender side until an acknowledgement was received.
Based on timeouts, an MQTT sender will retransmit application messages when an acknowledgement is missing.
\emph{QoS~2} ensures that a message is received exactly once, to avoid packet loss or processing of duplicates at the MQTT receiver side.
This requires a two-step acknowledgement process and more states at both sides.

To adapt MQTT to constrained networks which are based on low data rates and very small packet lengths such as in 802.15.4, MQTT-SN~\cite{mqttsn12} is specified.
Header complexity is reduced by replacing topic strings by topic IDs, to identify content.
In contrast to MQTT, MQTT-SN is able to run on top of UDP.
It still supports all QoS levels but does not inherit any reliability property from the transport layer.

\begin{table*}
\caption{Comparison of CoAP, MQTT, and ICN protocols. CoAP and MQTT support reliability only in confirmable mode (c) and QoS levels 1 and 2 (Q1, Q2).}
\label{tbl:coapmqttndn}
 \centering
\begin{tabular}{lcccccccc}
  \toprule
  & \multicolumn{5}{c}{\textbf{Current IoT Protocols}} & \multicolumn{3}{c}{\textbf{ICN Protocols}}
  \\
  \cmidrule(r){2-6}
  \cmidrule(l){7-9}
  & \multicolumn{3}{c}{CoAP~\cite{RFC-7252}} & MQTT~\cite{mqtt311} & MQTT-SN~\cite{mqttsn12} & NDN~\cite{jstp-nnc-09,zabjc-ndn-14} & I-Not~\cite{acim-ndnia-14} & HoPP~\cite{gksw-hrrpi-18}
  \\
  & PUT & GET & Observe & &
  \\
  \midrule
  Transport & UDP & UDP & UDP & TCP & UDP & n/a & n/a & n/a
  \\
  Pub/Sub & \xmark & \xmark & \cmark & \cmark & \cmark & \xmark & \xmark & \cmark
  \\
  Push & \cmark & \xmark & \cmark & \cmark & \cmark & \xmark & \cmark & \xmark
  \\
  Pull & \xmark & \cmark & \xmark & \xmark & \xmark & \cmark & \xmark & \cmark
  \\
  Flow Control & \xmark & \xmark & \xmark & \cmark & \xmark & \cmark & \xmark & \cmark
  \\
  Reliability & (c) & (c) & \xmark & (Q1, Q2) & (Q1, Q2) & \cmark  & \cmark & \cmark
  \\
  \bottomrule
\end{tabular}
\end{table*}

\begin{figure}	
 \includegraphics[width=0.90\columnwidth]{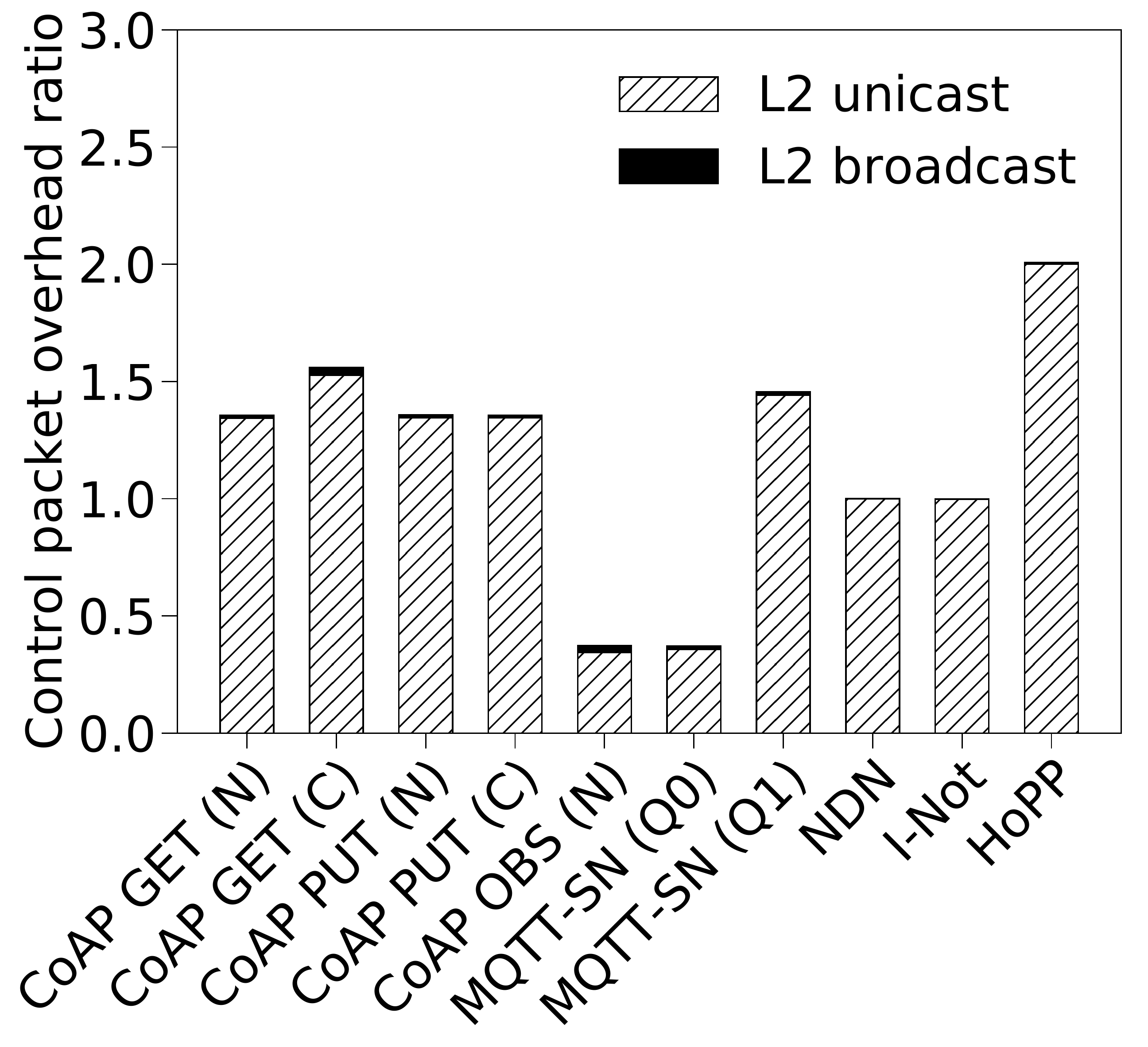}
    \caption{Relative protocol overhead under relaxed network conditions incl. topology control broadcasts.}
	 \label{fig:base_overhead}
\end{figure}

\subsection{ICN Protocols}

The core {\bf NDN} protocol~\cite{jstp-nnc-09,zabjc-ndn-14} combines name-based routing from TRIAD~\cite{gc-acrsi-01} and stateful forwarding from DONA~\cite{kccek-dona-07} to implement a request response scheme on the network layer.
Any consumer can request data that is subsequently delivered along a trail of reverse path forwarding states.
As an important feature, data will only be delivered to those who requested the data.
This means that data must be (individually) named at the Interest request and that yet unavailable data requires repeating Interests until the application receives the data.

The lack of push primitives in NDN triggered the idea of inverting the NDN semantic by placing data in an {\bf Interest Notification (I-Not)} which in turn gets acknowledged by the subsequent (empty) data packet. This idea was originally proposed in~\cite{acim-ndnia-14} and was since then criticized for its lack of \one caching support, \two flow control, and \three DDoS resilience. 

Several publish-subscribe extensions have been proposed for NDN (COPSS~\cite{cajfr-cecop-11}, PSync~\cite{zlw-sndsn-17}) to provide further decoupling of consumers and data sources. As COPSS relies on a persistent forwarding infrastructure and PSync on Interest broadcasting, both schemes do not satisfy the requirements of the constrained IoT.
Our lightweight IoT variant {\bf HoP and Pull (HoPP)}~\cite{gksw-hrrpi-18} provides a publish-subscribe system for constrained IoT deployments based on ICN/NDN principles.
A constrained IoT publisher announces a name towards a content proxy to trigger content requests and to replicate the data towards a content proxy (or broker).
Forwarding nodes on the path between publisher and content proxy hop-wise request content for this name by using common Interest and data messages.
A content subscriber  in  HoPP  behaves  almost  like  any  content requester  in  NDN and issues  a  regular  Interest  request  towards the content proxy CP.
However, in contrast to NDN \one a subscriber cannot extract  content  names  from  its  FIB,  since  FIBs  only contain PANINI default routes~\cite{swbw-lcnhp-16}, but uses application-specific topic tables instead; \two  it  does  not  expect  an  immediate reply, but issues Interests with extended lifetimes. HoPP enables rapid communication of unscheduled data events. It operates at a similar timescale as push protocols without actually pushing data.

\subsection{Protocol Comparison and Use Cases}

 Key properties of the three protocol families NDN, CoAP, and MQTT and  its variants are compared in Table \ref{tbl:coapmqttndn}. Specialized properties of the different approaches become apparent: Every protocol variant features distinct capabilities. Notably in the IoT, where TCP (aka generic MQTT) is unavailable, the pull-based NDN and NDN-HoPP are the only protocols admitting flow control and reliability as a generic service.  

 An additional mechanism for link recovery and retransmission has been brought to NDN with NDNLP~\cite{sz-nlpd-12}. Facing the lossy nature of low-power wireless links in the IoT, it may be tempting to deploy this additional protocol to enhance the overall reliability. However, common radio links like IEEE 802.15.4 already feature ARQs (Automatic Repeat-reQuests), and a network layer link would put a second acknowledgement to the air, which in turn would increase the omnipresent risk of interference. For this reason, we did neither deploy nor further investigate NDNLP in our further analyses. 

Figure \ref{fig:base_overhead} compares the control overhead for all protocol variants under consideration as obtained from experiments under relaxed network conditions at negligible interference. Aside from topology building and maintenance that are mainly broadcasts (marked in Fig.~\ref{fig:base_overhead}), common request protocols require one request per data item, whereas publish-subscribe schemes only require subscription notification per topic. As a pull protocol, HoPP requires requests and an additional message to advertise names.

\begin{figure}
	\resizebox{0.90\columnwidth}{!}{\input{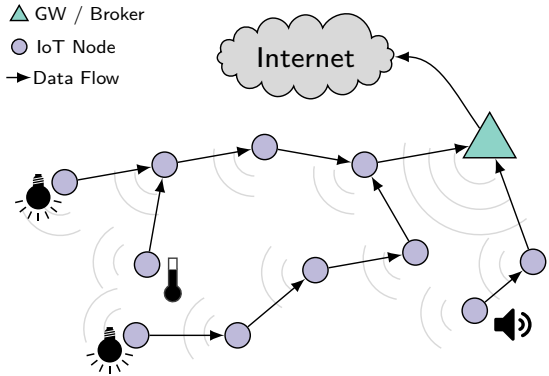}}     
    \caption{Use case scenario of a multi-hop IoT topology.}
	\label{fig:deployment_scenario1}
\end{figure}

Common IoT deployment use cases consist of stub networks as visualized in Figure~\ref{fig:deployment_scenario1} that may be single- or multi-hop. Traffic flows from or to the IoT edge nodes in three patterns: \one scheduled periodic sensor readings, \two unscheduled and uncoordinated data updates, or \three on demand notifications or alerting. It is worth noting that the different protocol properties (e.g., push versus pull versus pub-sub) can serve these alternating needs in a quite distinct manner.

\section{Implementation and Experimental Setup}
\label{sec:setup}

\paragraph{Software Platforms}
On the IoT nodes, all of our experiments are based on RIOT version \texttt{2018.01}.
To analyze CoAP, MQTT-SN, and NDN we use \texttt{gCoAP}, \texttt{Asymcute}, and \texttt{CCN-lite} respectively.
All three protocol implementations are part of the common RIOT release and thus reflect typical software components used in low-end IoT scenarios.

On the brokers or gateways, the testbed infrastructure deploys Linux systems.
To support MQTT broker and CoAP observe client as well as CoAP PUT server functionalities, we used \texttt{aiocoap} version~0.3 and \texttt{mosquitto.rsmb} version~1.3.0.2.
Both are popular open source implementations in this context.

\paragraph{Testbed}
We conduct our experiments in the FIT IoT-LAB testbed.
The hardware platform consists of typical class~2 devices~\cite{RFC-7228} and features an ARM Cortex-M3 MCU with 64~kB of RAM and 512~kB of ROM.
Each device is equipped with an Atmel AT86RF231~\cite{a-lptzi-09} transceiver to operate on the IEEE 802.15.4 radio.
The gateway runs on a Cortex-A8 node, which is more powerful than the M3 edge nodes.

The testbed provides access to several sites with varying properties.
We perform our experiments on two sites, to analyze single-hop as well as multi-hop scenarios.
\begin{description}
    \item[Single-hop topology] The \emph{Paris} site consists of approximately 70~nodes, which are within the same radio range.
        We choose two arbitrary nodes  and run all single-hop experiments on them. One node is a
        content producer, the other node acts as consumer (gateway/broker).

    \item[Multi-hop topology] The \emph{Grenoble} site consists of approximately 350 nodes spread evenly in the Inria Grenoble building.
      We choose 50~M3 nodes (low-end IoT device) and one A8 node (gateway/broker) arbitrarily and
      run all multi-hop experiments on them. All low-end devices operate as content producers.
      In our CoAP and MQTT experiments, we use RPL to build and maintain the routing topology across
      all nodes. In our NDN-based experiments, we build tree topologies analogously as HoPP does.
      In any case, we ensure that all protocols use the same routing topology for comparison.
      Typical path length are four to five hops.
\end{description}

\paragraph{Scenarios and Parameters}
We align all experiments with respect to the configurations of retransmissions and timeouts to ensure comparability among protocols.
All protocols employ the same retransmission strategy: In case of failures, each node waits 2~seconds before retransmitting the original application or control data.
For NDN, HoPP and I-Not, retransmissions are performed hop-by-hop, while CoAP and MQTT perform them end-to-end.
At most 4~retransmissions will occur for each data.
Interest lifetimes are configured to 10~seconds for NDN based protocols to limit PIT memory consumption.
We repeat each experiment 1,000 times.

To accommodate all 50 nodes in the routing topology, the FIB size was adjusted accordingly  on each IoT node.
For CoAP and MQTT, this translates in our IPv6 scenario to a FIB size of 50 entries with roughly 32 bytes each (\texttt{sizeof(destination)} + \texttt{sizeof(next-hop)}).
In our NDN scenarios, each node owns a unique prefix of the form /$\rho_i$ with a length of \texttt{24}
bytes.
The next-hop face of each FIB entry points to the \texttt{8}-byte IEEE 802.15.4 link-layer address.
In total, this setup yields comparable size requirements for all scenarios.

In the NDN scenarios, we use unique content names prefixed by /$\rho_i$
with incremental local packet counters. CoAP works without unique names but uses common URIs.
The MQTT-SN protocols register a common topic name, similar to CoAP, and publish under a unique
topic~ID  thereafter. In all scenarios, the data is of the same JSON format consisting of a unique
identifier and a sensor value attribute. These short messages can be accommodated by the link layer and do not require fragmentation. It is noteworthy that we neither applied header compression in the IP~\cite{RFC-7400} nor in the NDN world~\cite{draft-icnrg-ccnlowpan}.   

\section{Evaluation and Results}
\label{sec:eval}

\subsection{Analyses and Metrics}

The objective of this work is to quantify the efficiency and utility of the considered protocols in real deployment scenarios. With this in mind, we want to shed light on resource consumption and the operational properties of data dissemination from different angles and in the different deployment use cases.

In detail, we analyze the \emph{memory consumption} on nodes, the effective \emph{network
utilization} by control and data traffic including \emph{protocol overhead} and \emph{link stress}
caused by retransmissions. The actual performance of data transmission is measured in \emph{data
loss}, \emph{goodput}, and \emph{content arrival time} which represents the delay between issuing a
transaction and data arrival at the sink.
 Here, we use the term
\emph{time to completion} interchangeably. We also consider the \emph{data
flows} and its \emph{energy consumption}.
These multi-sided  analyses are performed on complete packet traces which we recorded from the
different experiments, and a monitoring of the system state at participating nodes.

Security measures largely differ between the IP and the ICN world. DTLS \cite{RFC-6347} provides privacy and integrity for UDP datagrams within sessions based on pre-established private keys. NDN authenticates data chunks between arbitrary endpoints without the need for session state. Canonically, asymmetric signatures are attached to data chunks in NDN, but since the complexity of asymmetric crypto exceeds the capabilities of constrained nodes, keyed-hash message authentication code (HMAC) can also be applied. The use of HMAC likewise relies on pre-established keys. 

In both worlds,  security extensions add message and processing overhead, but do not change the overall behavior of the protocols. For this reason, we compare security overheads in separate micro-benchmarks and perform the remaining experiments without applying the corresponding security measures.

We do not consider network congestion from external cross-traffic in this work. However,  each individual 
 transmission experiences self-induced background traffic from the experiment that differs for varying
request/publish intervals and jitter. On average, this side-traffic is constant per
experimental run.

\subsection{Protocol Stack Sizes}

\begin{figure}
     \includegraphics[width=0.50\textwidth]{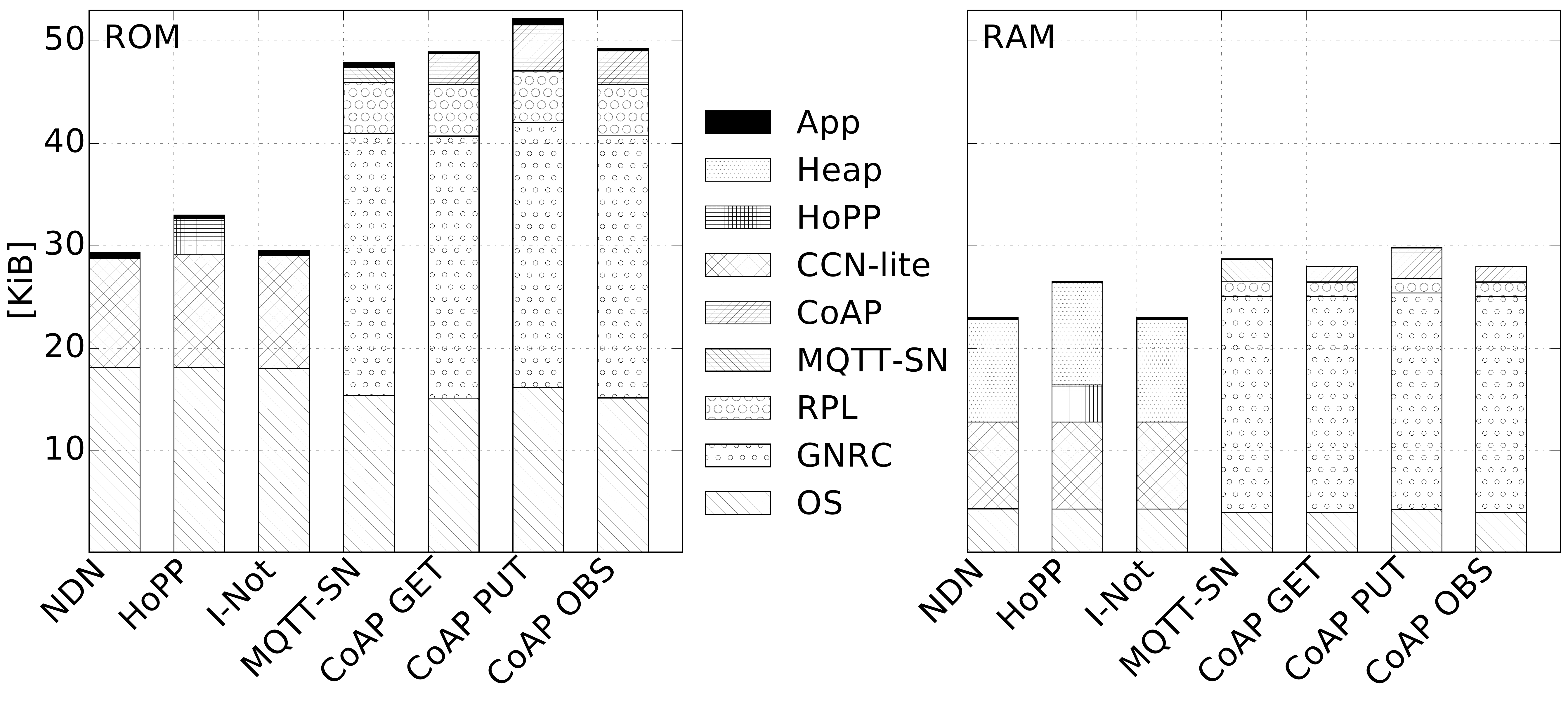}     
    \caption{Resource consumption of ROM (left) and RAM (right) for the different software stacks.}
      \label{fig:stacksizes}
\end{figure}

Largely differing properties and complexities of the protocol variants under test lead to seven distinct software stacks. Nodal memory consumption for these different protocol stacks are depicted in Figure \ref{fig:stacksizes}. We differentiate the  protocol layers in place to disclose the details.

Main memory is the scarcest resource in the IoT. While  protocols require OS support of 4,060~B (MQTT-SN) -- 4,400~B (NDN)  kernels, NDN admits the leanest stack of  8,700~B consumed by CCN-lite. All IP protocol stacks are significantly larger and approximately triple the size of CCN-lite. On the overall, about 30~KiB are needed to host IP protocols, leaving only a few dozen KiBs for the application on  typical constrained nodes.
 All ICN protocols provide a Content Store (CS) of  10,240~B on the heap, which is the price of in-network caching. It should be noted that the GNRC network stack contributes a packet buffer to both, the IP and the ICN world that is also used for retransmissions \cite{lkhpg-cwemr-18}. 
Program sizes of NDN protocols are much smaller and consume about 40~\% less ROM. The operating system support varies with protocol requirements on the highly modular RIOT OS platform.

\subsection{Security Overheads}

Many use cases of the IoT rely on integrity and authenticity of the collected data. Security extensions of the communication protocols are requested to ensure those properties at costs which we are now evaluating. For our micro-benchmarks of the IP world, we fixed the  scenario of a DTLS session established between two nodes. We quantify the messaging overhead obtained from a single session establishment and the packet overhead as a function of data transactions---the request/response-guided transfer of a data unit. We also recorded the CPU expenses at the content producer and consumer per transaction.    

The most comparable scenario for NDN consists of HMAC-based authentication of data using SHA256 per chunk. For quantifying the overheads in data packets, we chose two common sizes of the KeyLocatorTLV: 16B and 32B. 

\begin{figure}
    \includegraphics[width=.5\textwidth]{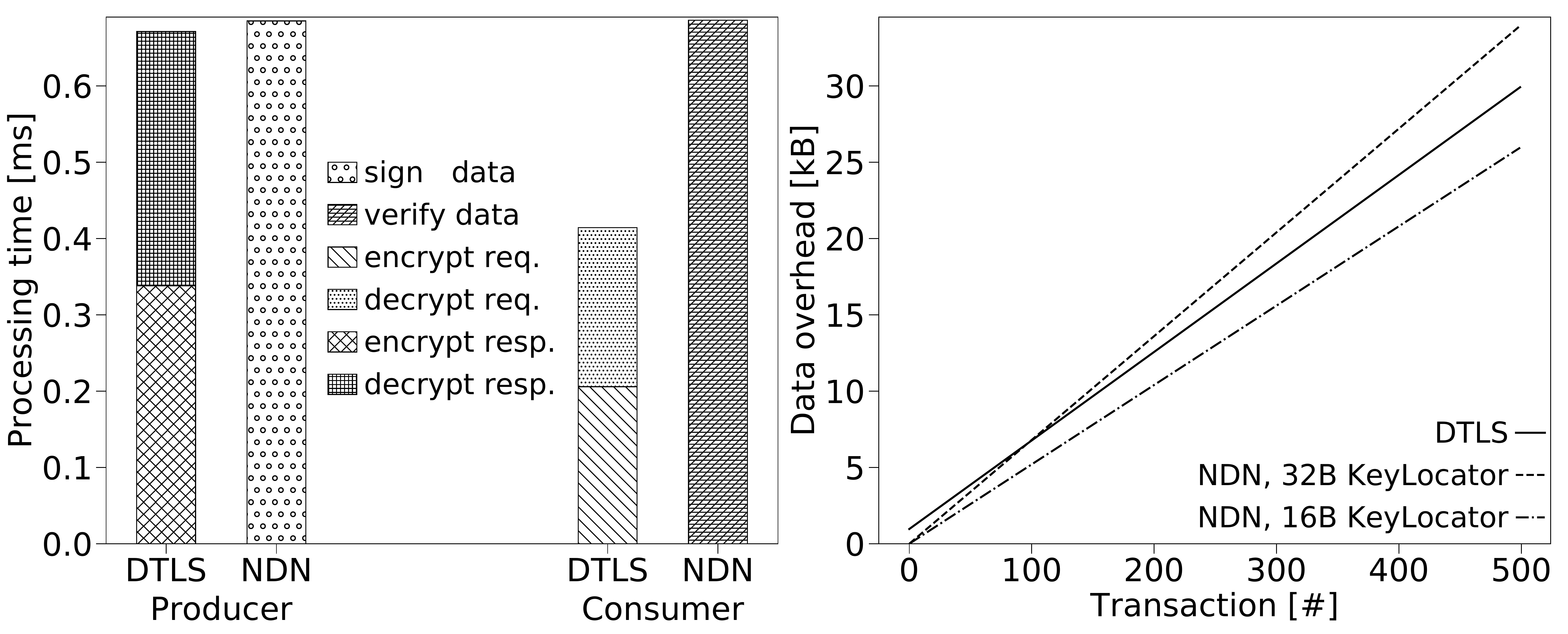}
    \caption{Security overheads---CPU  consumption (left) and data overhead (right) per content transaction for IP/DTLS and
        NDN/HMAC.}
    \label{fig:security_overhead}
\end{figure}

Figure~\ref{fig:security_overhead} visualizes the results of our security benchmarks performed on the IoT-LAB M3 nodes. While message overheads for NDN are similar or better (for 16B KeyLocatorTLV), DTLS data verification can be performed at two-thirds of the NDN costs. It should be noted, though, that the different security models of DTLS and NDN make comparisons difficult. While DTLS operates within an established session that is strictly bound to endpoints, the content of NDN can be replicated between varying nodes. In particular, the NDN approach is robust under mobility and network changes, whereas DTLS would require to re-establish sessions in many cases at significant cost. Conversely, only DTLS encrypts transport payloads and thereby contributes data privacy.

\subsection{ Single-Hop with Scheduled Publishing}

\begin{figure*}
   \subfigure[Packet loss at 50 ms publishing interval]{\includegraphics[width=0.32\textwidth]{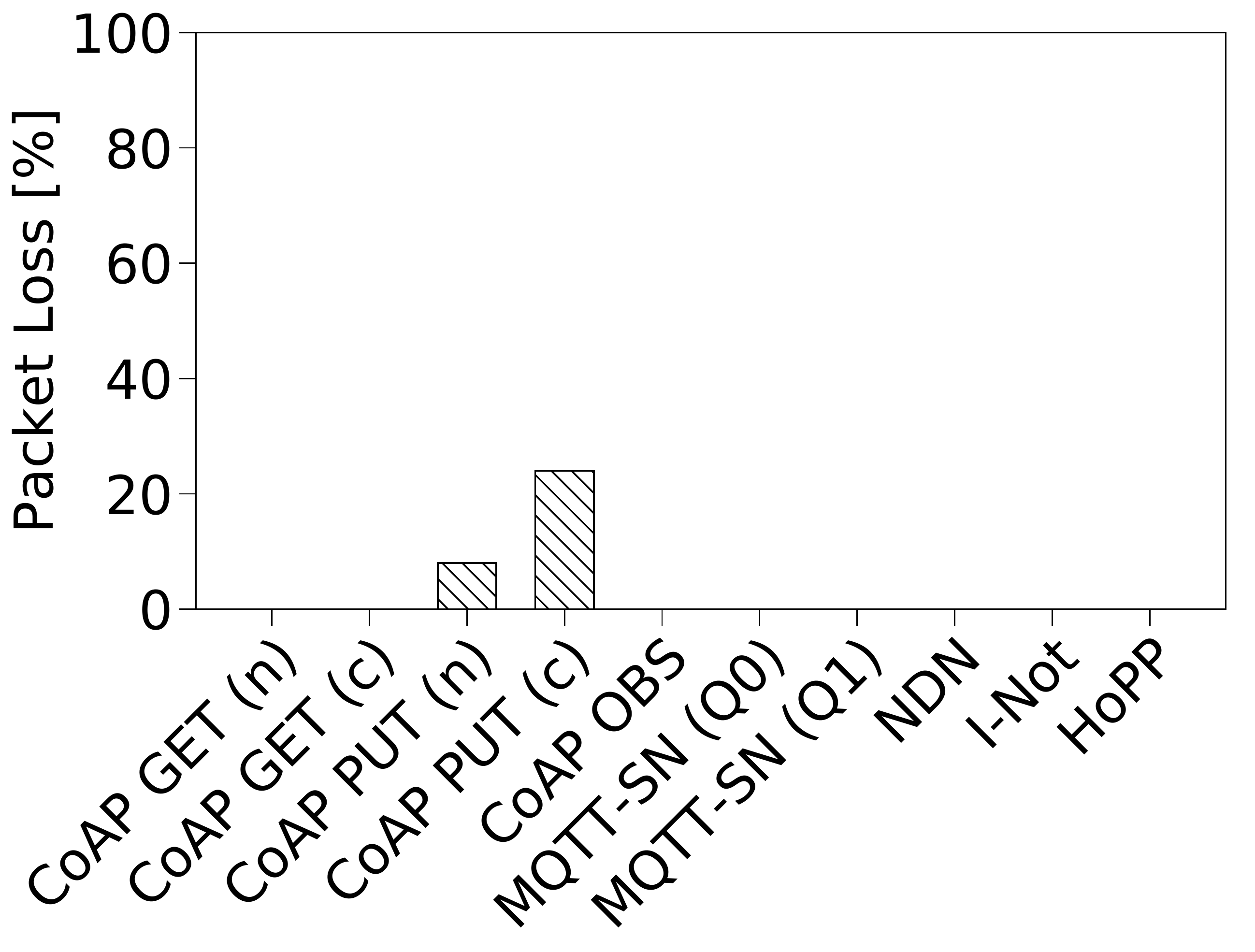} \label{fig:single_loss}}
      \subfigure[Push at 50 ms publishing interval]{\includegraphics[width=0.32\textwidth]{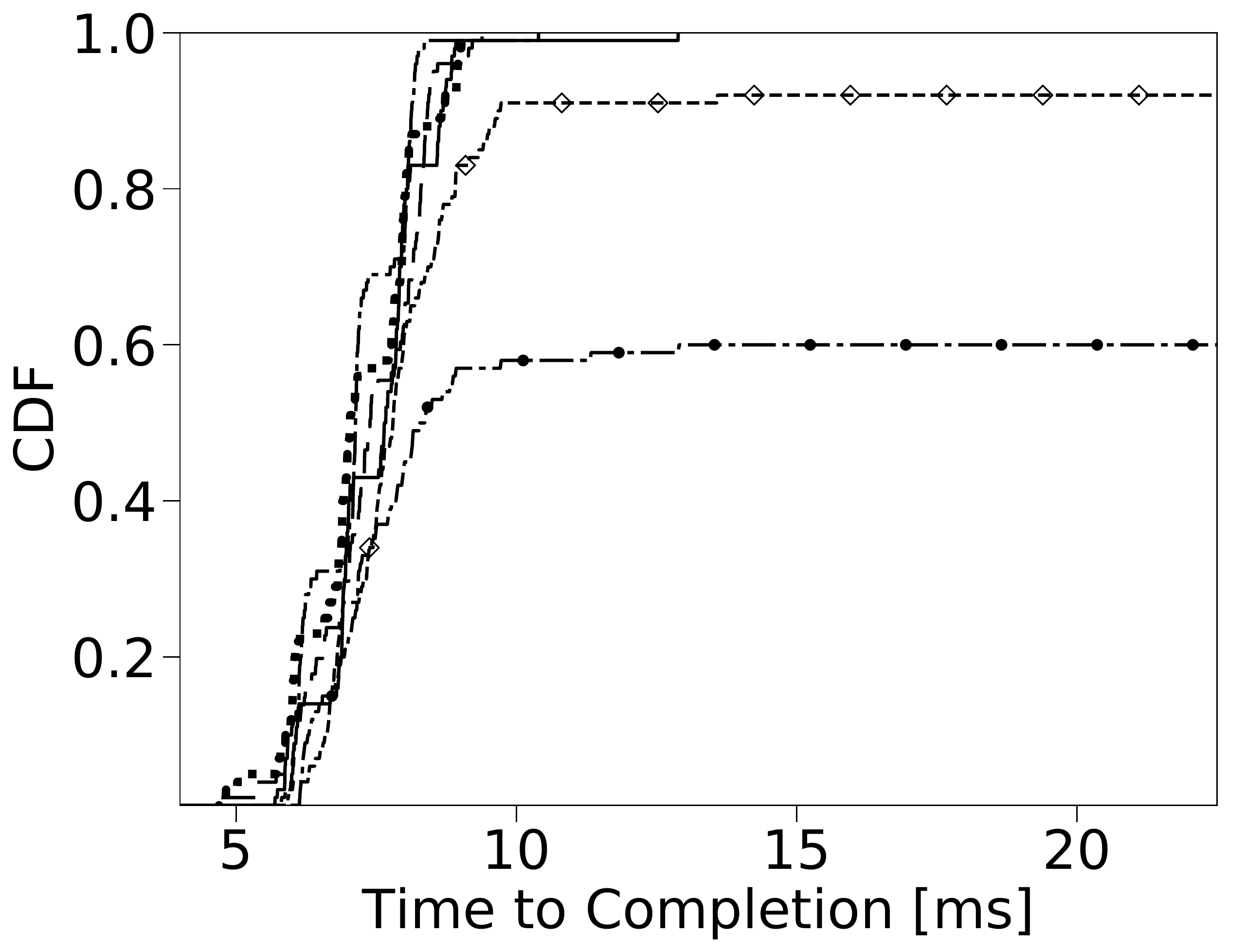} \label{fig:ttc_singlepush_50ms}}
     \subfigure[Push at 5 s publishing interval]{\includegraphics[width=0.32\textwidth]{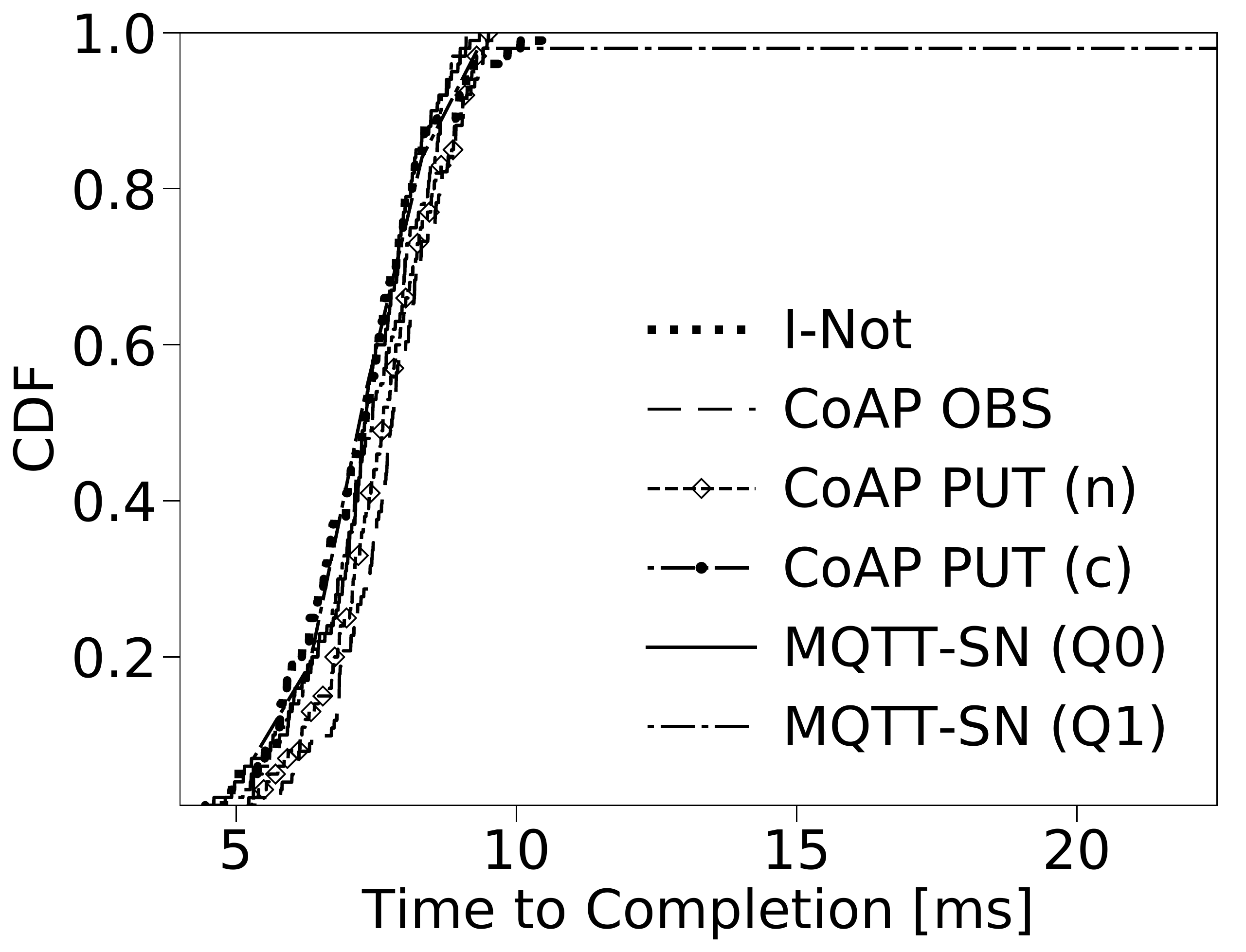} \label{fig:ttc_singlepush_5s}}
	\subfigure[L3 retransmissions at 50 ms  interval]{\includegraphics[width=0.32\textwidth]{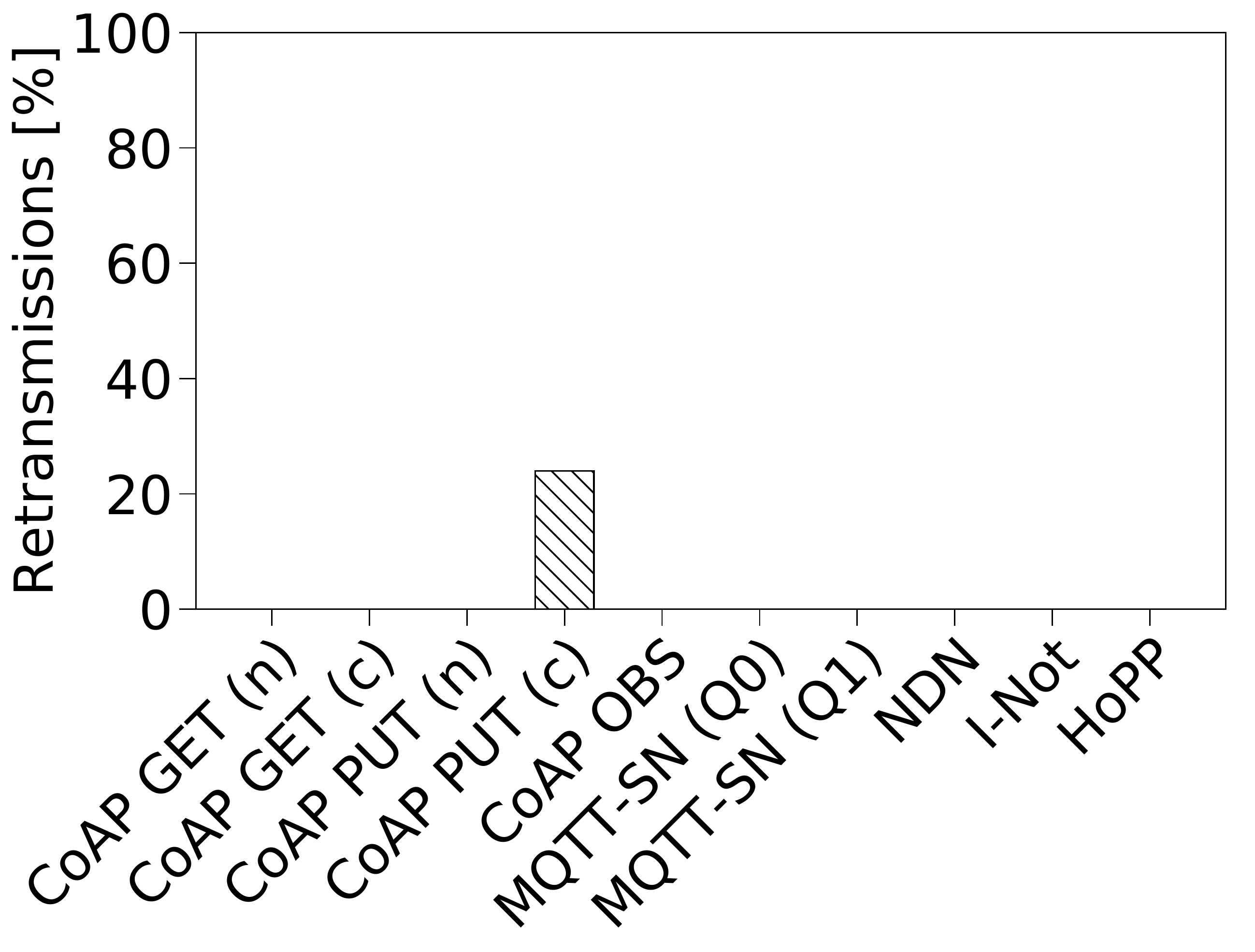} \label{fig:single_retrans}}
      \subfigure[Pull at 50 ms publishing interval]{\includegraphics[width=0.32\textwidth]{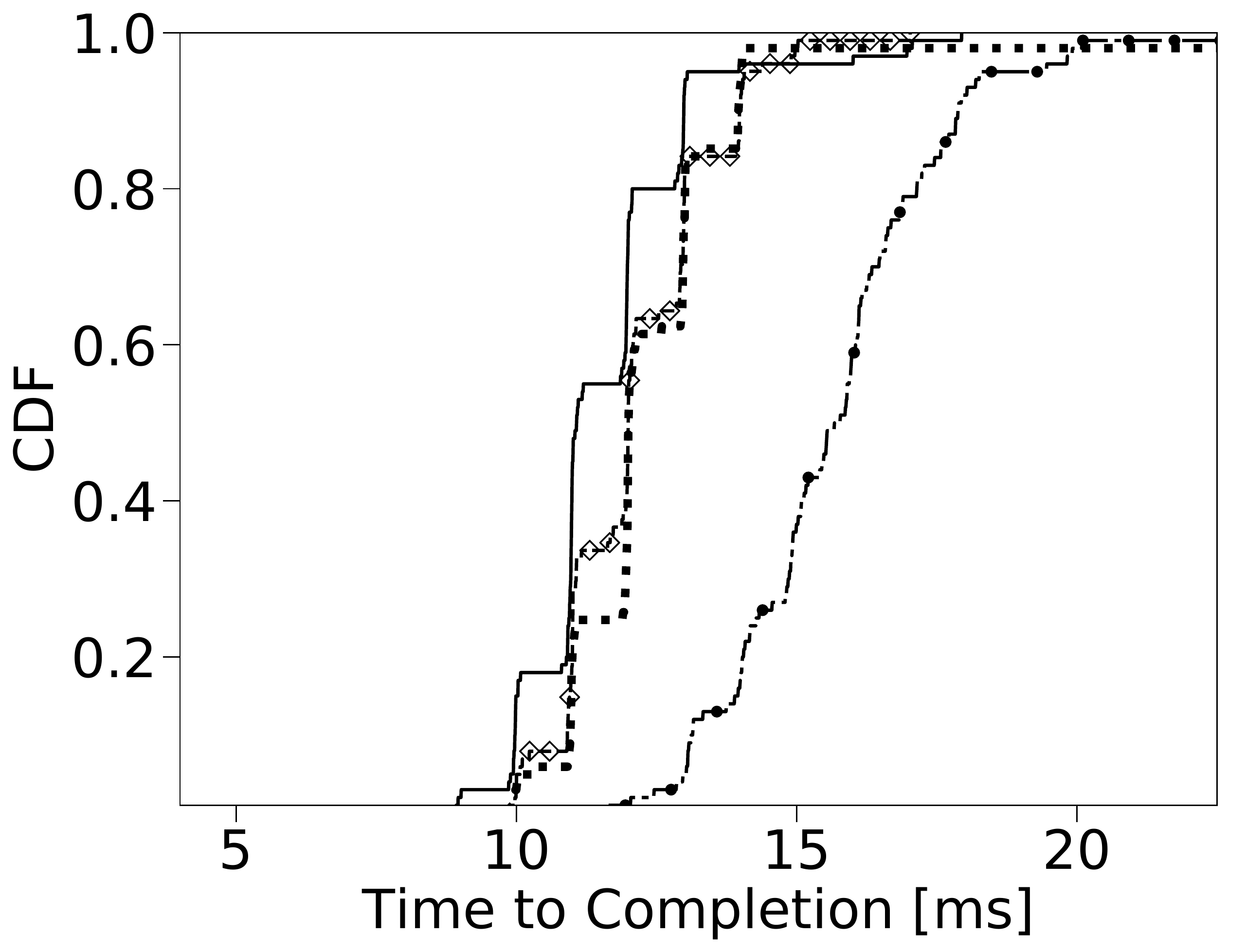} \label{fig:ttc_singlepull_50ms}}
 \subfigure[Pull at 5 s publishing interval]{\includegraphics[width=0.32\textwidth]{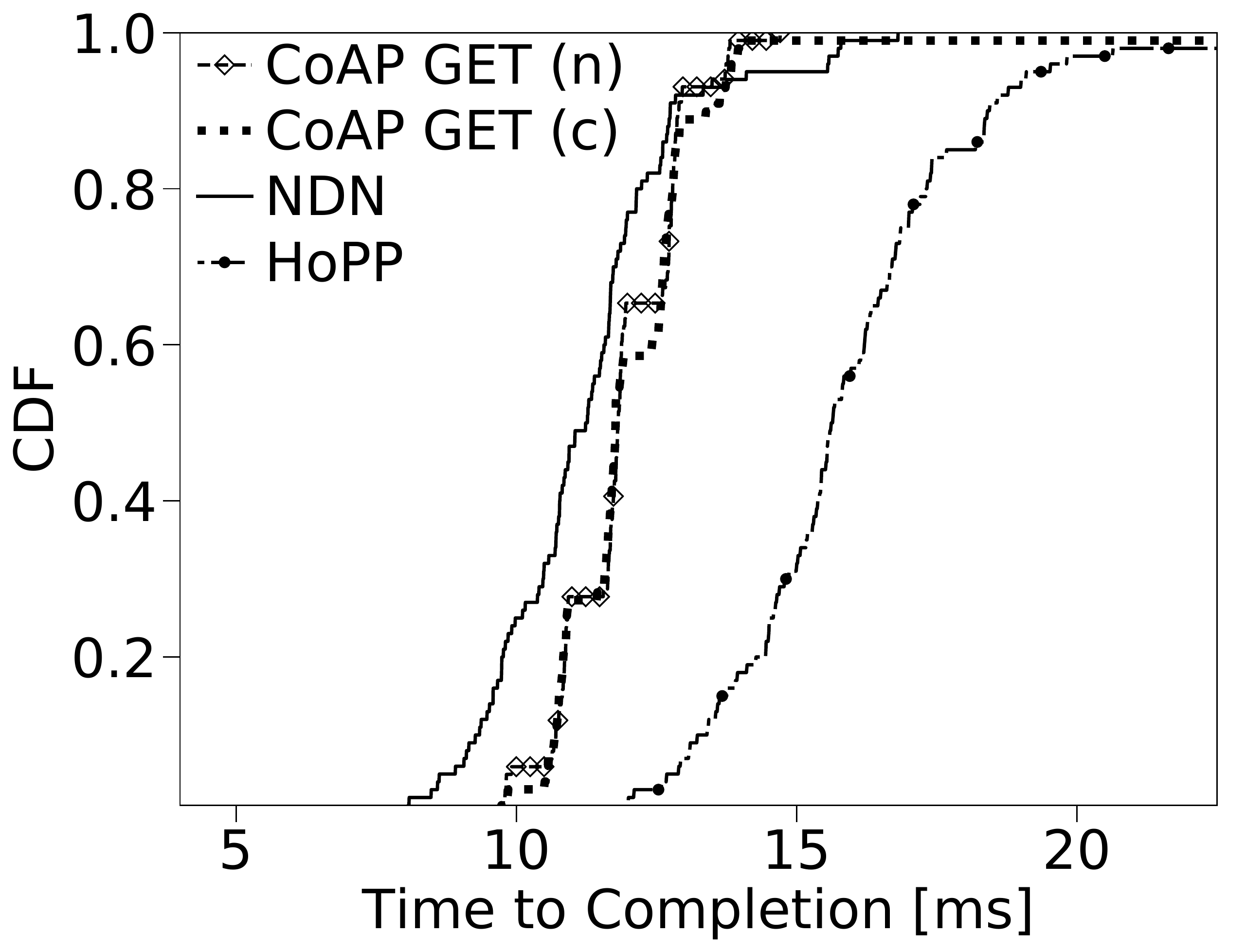}\label{fig:ttc_singlepull_5s}}
    \caption{Time to content arrival for scheduled publishing in a single-hop topology at different  intervals.}
      \label{fig:single-scheduled}
\end{figure*}

Protocol performances are first evaluated in a single-hop topology at the Paris testbed with periodic content publishing every $50\ ms$ and $5\ s$. Content is pushed or requested accordingly. Figure \ref{fig:single-scheduled} displays the results for protocol reliability and temporal performance. As an overall trend, it is apparent that push-oriented protocols operate faster, but less reliable. 

For the rather relaxed scenario of publishing every $5\ s$, we see the protocol families in rough agreement. Push-based protocols require an average of $7\ ms$ (Fig.~\ref{fig:ttc_singlepush_5s}) for data delivery, pull-based protocols take  $11~ms$ (Fig. \ref{fig:ttc_singlepull_5s}), with the exception of HoPP which is slightly slower on this short timescale due to its three-way handshake.

The publishing interval of $50\ ms$ puts some protocols under stress, even though  IEEE 802.15.4  practically limits  transmission  only below an interval of  $10\ ms$. The performance of CoAP PUT significantly degrades (Fig. \ref{fig:ttc_singlepush_50ms}), leaving the unconfirmed messaging at a total data loss of 6~\% (Fig. \ref{fig:single_loss}). The PUT of Confirmable CoAP  instead initiates 26 \% retransmissions (Fig. \ref{fig:single_retrans}) which increase delays up to a factor of five. Confirmable CoAP does complete data delivery at $42\ ms$ (Fig. \ref{fig:ttc_singlepush_50ms} is clipped for visibility). It should be noted, though, that retransmissions on the data link layer  are present for all protocols and are reflected by the staircase patterns.  We do not measure these fast repeats ($\leq 10\ ms $)  in this work, but refer to our previous study \cite{kgshw-nnmam-17} for further details.

\subsection{Single-Hop~with~Unscheduled~Publishing}

\begin{figure*}
     \center
     \subfigure[Control Overhead for polling unscheduled content]{\includegraphics[width=0.9\columnwidth]{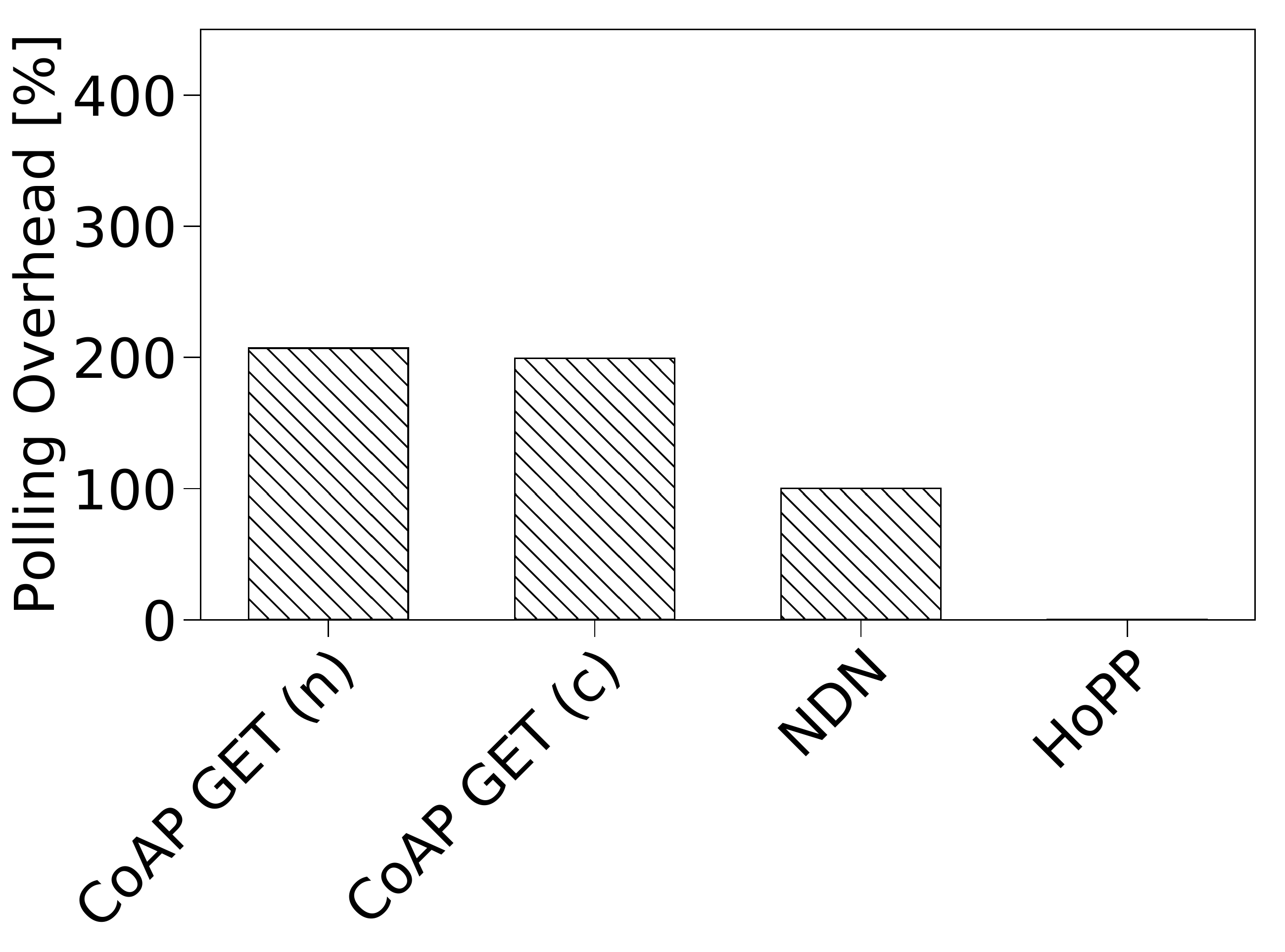} \label{fig:coh_unscheduled}}
	  \quad
      \subfigure[Time to unscheduled content arrival]{\includegraphics[width=0.9\columnwidth]{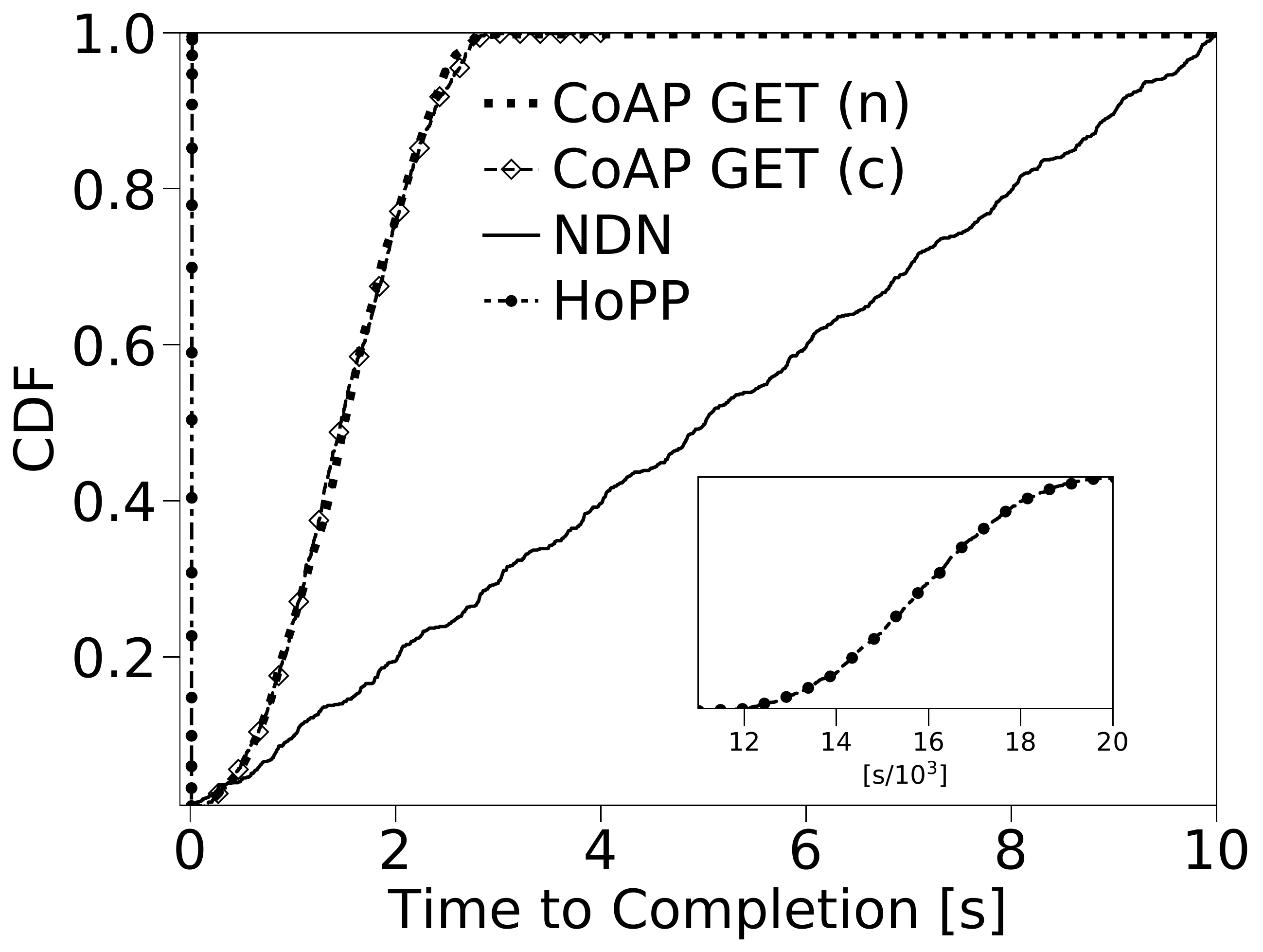} \label{fig:ttc_unscheduled_2000}}
	\caption{Pull protocol performance at random publishing in [1s $\dots$ 3s].}
      \label{fig:single_unscheduled}
\end{figure*}

Our next experiments address the common IoT use case of publishing data at irregular intervals. This is the typical pattern for observing third party actions (\eg light switching), or largely uncoordinated sensing environments. Push-based protocols naturally serve these application needs. We quantify the behavior of the request-based protocols in practice and chose the moderate setting of publishing content every two seconds on average. Publishing is uniformly distributed in the interval of [1\,s $\ldots$ 3\,s]. The  protocols CoAP and NDN request the content periodically every second so that updates are not lost.

Figure \ref{fig:ttc_unscheduled_2000} visualizes content delivery times for all request-oriented protocols. CoAP GET and NDN now operate on a timescale of seconds, while HoPP continues to complete in the unaltered range of $15\ ms$ without additional protocol operations -- the unsurprising outcome of content triggers built into HoPP. CoAP requests content using a common name with the result of likely duplicate content transmissions. On average, CoAP needs two requests to retrieve fresh content with the expected average delay of $\approx 2\ s$ and a corresponding  polling overhead of 200 \% (Fig. \ref{fig:single_unscheduled}). In contrast, NDN admits lower overhead, as Interests are locally managed at the PIT and only retransmitted after state timeout.

However, issuing Interests at a higher rate than content arrival leads to an accumulation of open states in the PIT. As resources on the constrained nodes are tightly bound, the PIT limits are quickly reached and can be only met by either {\em discarding} newly arriving Interests, or by {\em overwriting pending Interest state}. Both countermeasures delay content delivery, as can be seen from Figure \ref{fig:ttc_unscheduled_2000}. In detail, the time to content delivery of NDN stretches over various PIT combinations up to the final PIT timeout of $10\ s$. It is noteworthy that PIT overflow in these experiments appears for available content  that is ready for delivery  via valid routes. NDN protocol extensions such as NACKs would neither help nor should be triggered, since Interest retransmissions act as counter measures to packet loss or timeouts due to wireless link degradations. Consequently, the quantitative impairment of packet delivery tightly depends on the scenario and can  lead to significant data loss in the constrained IoT, as well. 

These experiments shed again light on the  trade-off between memory and network performance in the NDN stateful forwarding regime as has been first identified in \cite{wsv-bipmc-12} and recently discussed in the IoT context \cite{sblwy-ndnti-16}.

\subsection{Multi-Hop Topologies}

We  now consider the more delicate use case of mixed communication in multi-hop topologies: 50 nodes exchange  content that is published every 5 or 30 seconds  in an uncoordinated manner. Repeated experiments were performed on the Grenoble testbed with tree topologies of routing depths varying from four to six hops. 

\begin{figure*}
     \centering
      \subfigure[Push at 5 s publishing interval]{\includegraphics[width=0.9\columnwidth]{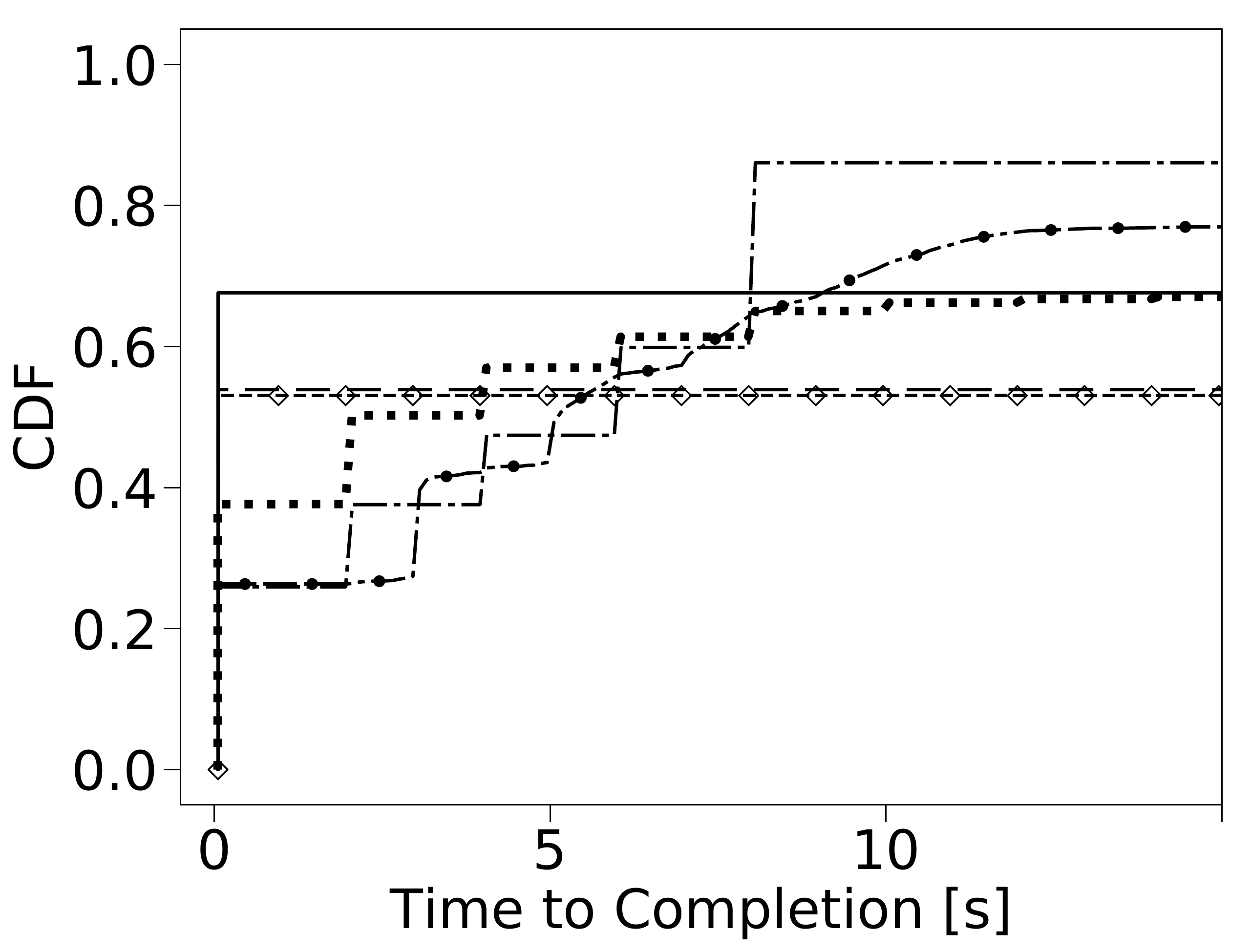} \label{fig:ttc_multipush_5s}}%
		\quad%
	    \subfigure[Push at 30 s publishing interval]{\includegraphics[width=0.9\columnwidth]{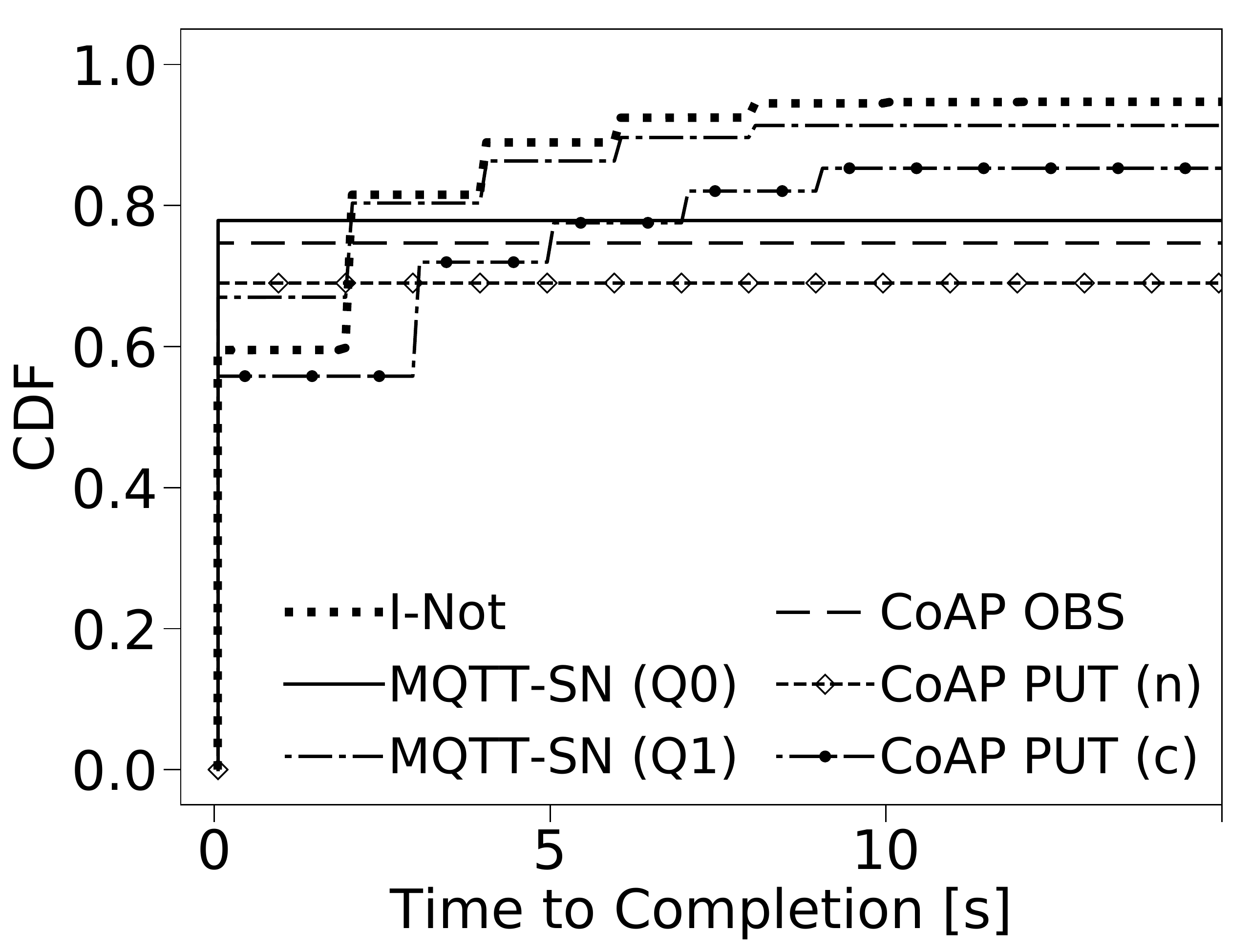} \label{fig:ttc_multipush_30s}}
	  \subfigure[Pull at 5 s publishing interval]{\includegraphics[width=0.9\columnwidth]{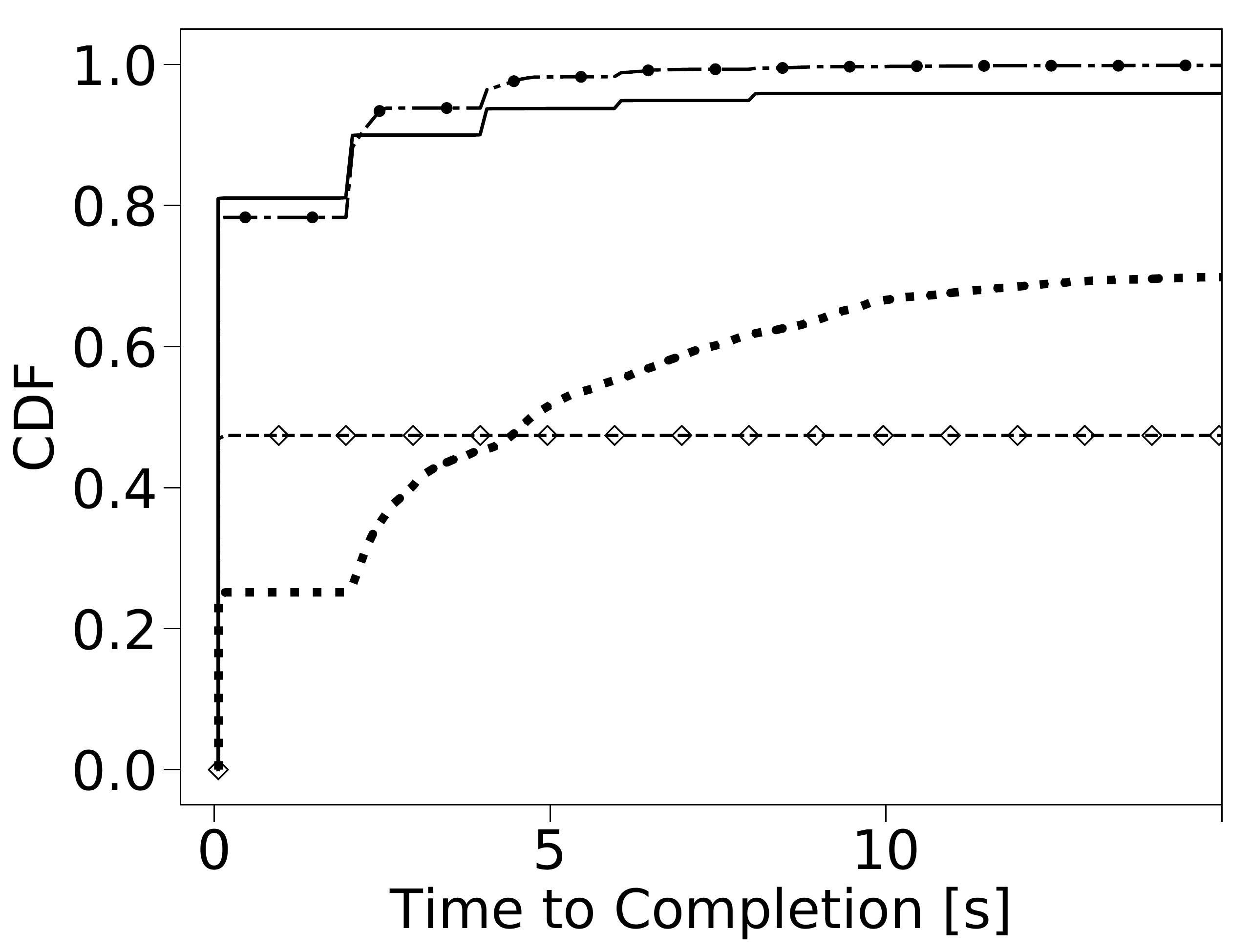} \label{fig:ttc_multipull_5s}}
		\quad%
        \subfigure[Pull at 30 s publishing interval]{\includegraphics[width=0.9\columnwidth]{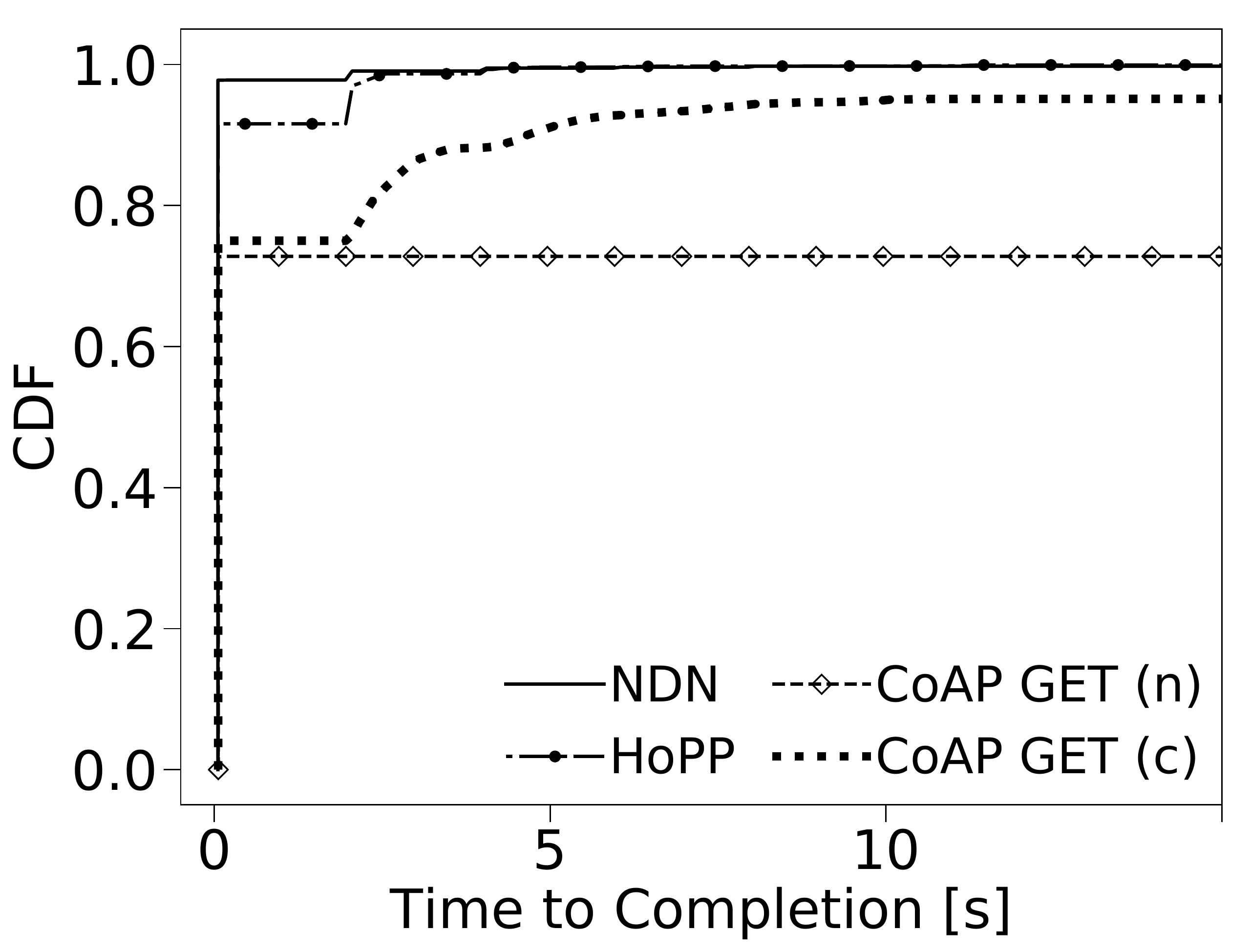}\label{fig:ttc_multipull_30s}}
    \caption{Time to content arrival in multi-hop topologies of 50 nodes.}
      \label{fig:multihop_scheduled}
\end{figure*}

First, we examine the temporal distributions from content publishing to arrival in analogy to the single-hop cases. Figure \ref{fig:multihop_scheduled} combines the results for push and pull protocols, as well as both publishing rates. The overall results reveal a much slower and less reliable protocol behavior than could be expected from the single-hop values in Figure \ref{fig:single-scheduled}. Graphs reflect the common experience in low power multi-hop environments that interferences and individual error probabilities accumulate in a destructive manner.

Push and pull protocols now operate on similar time scales in the absence of considerable disturbances, while events of strong interference and packet corruption on the air lead to large retransmission delays and loss. Protocol retransmissions with an interval of 2 seconds are clearly reflected by the staircase patterns in the respective CDF. Most notably, the `reliable' variants of CoAP PUT (c) and GET (c) fail to always transfer the content, but remain unsuccessful in a range between 5 \% (at 30~s publishing) and  30~\% (at 5~s). Even though confirmable CoAP operates more reliably than the unreliable versions OBS and PUT/GET (n), the failure rates indicate a quite unsatisfactory protocol behavior. A similarly unsuccessful performance must be observed for the NDN push variant Interest Notification.
In contrast, the reliable MQTT (Q1) successfully transfers its data in 90--95~\% of all cases, thereby heavily relying on retransmissions as we will see in the course of the further analysis.

The performance of NDN shows decent results both in promptness and reliability, even though 5 \% of data chunks remain lost in the fast publishing scenario (5~s). The only protocol that delivers reasonably fast at full reliability is the NDN variant HoPP. Below we will see that this happens with the least retransmissions and in evenly balanced flows. In a way, this result is not surprising as HoPP is optimized for IoT demands  and the only protocol that balances data transmissions per hop. It is the common experience in the low power wireless that  link qualities vary quickly and largely.

\begin{figure*}
     \includegraphics[width=1.0\textwidth]{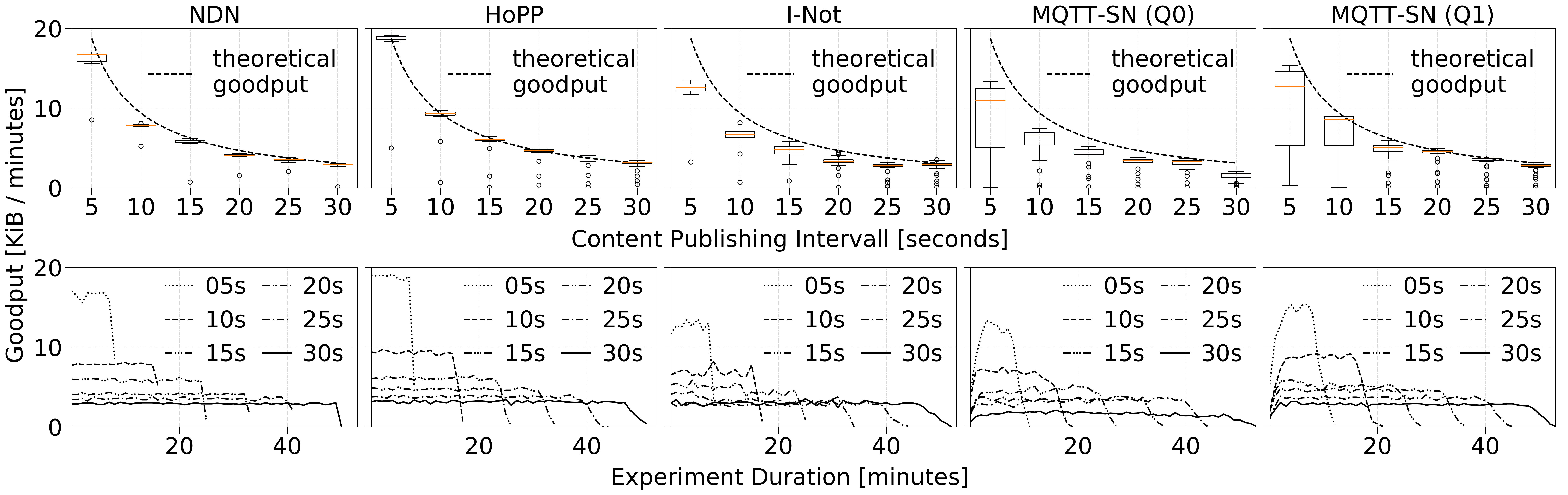}     
    \caption{Goodput summary and  evolution  for the NDN and MQTT protocols at different publishing intervals.}
      \label{fig:ndn_goodput}
\end{figure*}

\begin{figure*}
     \includegraphics[width=1.0\textwidth]{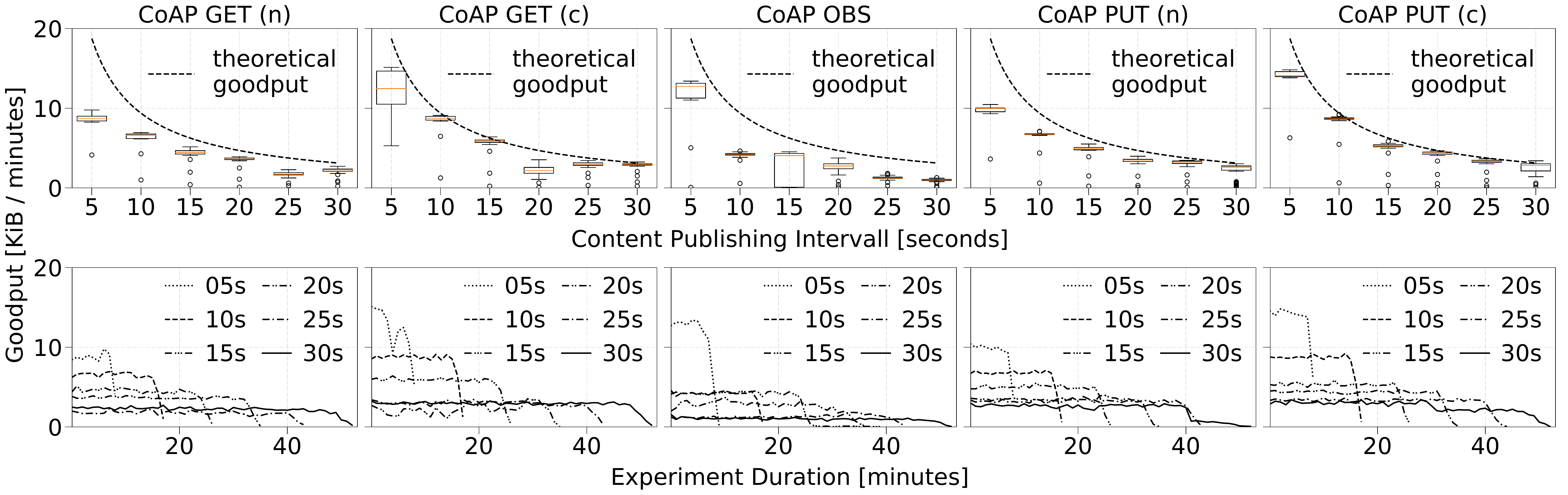}     
    \caption{Goodput summary and  evolution  for the CoAP protocols at different publishing intervals.}
      \label{fig:ip_goodput}
\end{figure*}

Second, we evaluate the effective data goodput and flow analysis of the different protocols during content publishing experiments. In Figures \ref{fig:ndn_goodput} and  \ref{fig:ip_goodput}, we summarize the results for the variants of NDN, MQTT, and CoAP respectively. We display the different experimental results of the data goodput in box plots and compare to the theoretical optimum (lines). Time series of data goodput are further revealing the flow behavior as displayed in the lower row of these figures. 

Clearly, HoPP admits the most evenly balanced flows and shows nearly optimal goodput values, closely followed by NDN. All other flow performances fluctuate with some tendency of instability when approaching its full transmission speed. Some IP-based flows in MQTT and CoAP drop to lower delivery rates which is dominantly caused by slow repeated end-to-end retransmission. Multi-hop retransmissions in this error-prone regime tend to cause additional interferences and accumulate transmission errors. As a consequence, protocols operate at reduced efficiency -- in some cases protocol performance drops down to 50~\% (\eg CoAP GET (n) and CoAP OBS in Fig.  \ref{fig:ip_goodput}). Interest Notification which is not capable of content caching, does not outperform the IP protocols. The overall results show that the absence of flow control as in UDP/IP--based protocols and in the I-Not variant of NDN make protocols fragile.
Hop-wise retransmission management as applicable in  NDN and HoPP re-balances flows and explicitly demonstrates its benefits for the IoT instead. 

\begin{figure*}
     \includegraphics[width=1.0\textwidth]{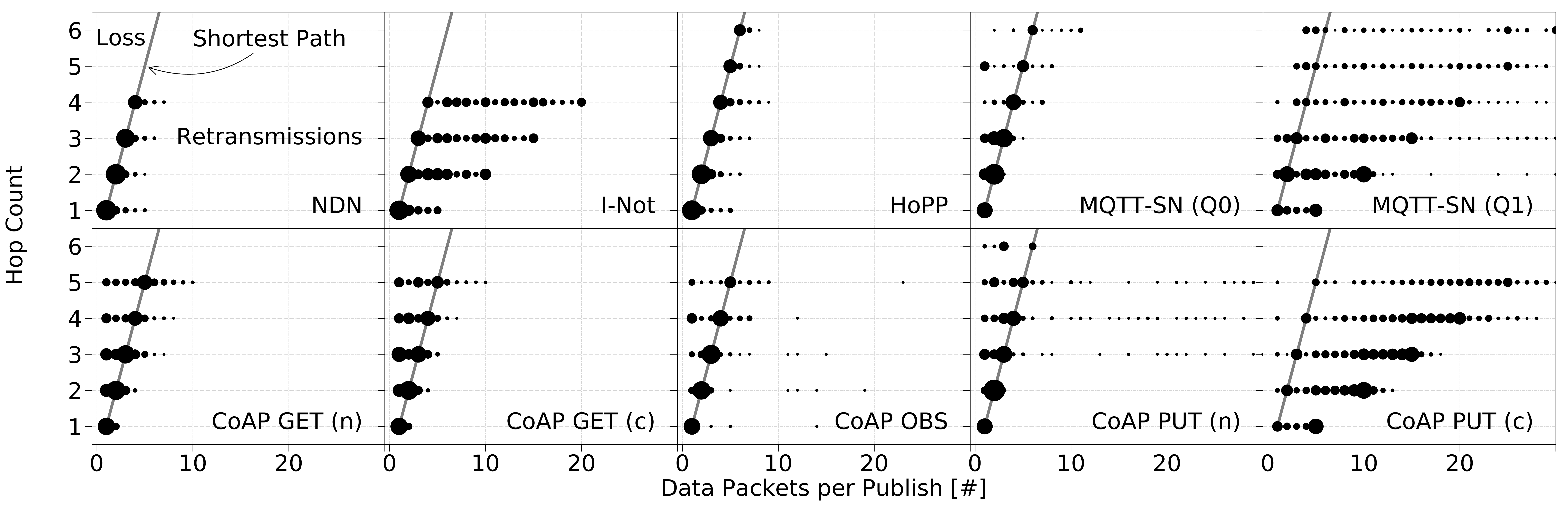}     
    \caption{Link traversal vers. shortest path for a 15 s publishing interval. The scatterplots reveal the link stress with dot sizes proportional to event multiplicity.}
      \label{fig:link_stress}
\end{figure*}

Our next evaluation focuses on the link utilization. We measure all individual paths that each unique data packet traveled on its destination from source to sink and contrast the results with the corresponding shortest possible path. Results are visualized as scatterplots in Figure \ref{fig:link_stress}. Each dot represents one or several events, the dot size is drawn proportionally to  event multiplicities. Solid lines indicate the shortest paths, while events left of the line represent failures (traversal shorter than the shortest path). Right of the solid diagonal retransmissions are counted.

The ideal protocol performance is situated on the diagonal line with all data traversing each link only once on the shortest path. This ideal behavior is most closely approximated by the NDN core and the NDN HoPP protocols. A largely contrasting performance can be seen from the reliable IP protocols MQTT (Q1) and CoAP PUT~(c) which admit huge numbers of retransmissions. This also holds for the NDN Interest Notification protocol which cuts out the NDN feature strength by inverting its semantic.

Unreliable IP-based protocols show very large loss multiplicities and only a few retransmissions which are initiated by reacting to link-layer failures. This corresponds to the reduced success rate already observed in the previous evaluations. Apparently, all protocols that follow an end-to-end path semantic (including I-Not) are forced to struggle against the unpredictable nature of intermediate links---either by voluminous packet retransmissions or significant packet loss. 

In our final experimental comparison between the protocols, we evaluate the individual energy consumption per node as a function of time. Since the energy demand of a protocol is largely dominated by its radio transfer of packets, we focus our measurements on `bytes in the air', i.e., the  IEEE 802.15.4 transmission and reception of packets on each individual node.
Power consumption levels for transmitting, receiving and idling are obtained from the Atmel AT86RF231 data sheet and we calculate the actual energy from measuring the radio operation time in the respective device state. 

\begin{figure*}
     \includegraphics[width=1.0\textwidth]{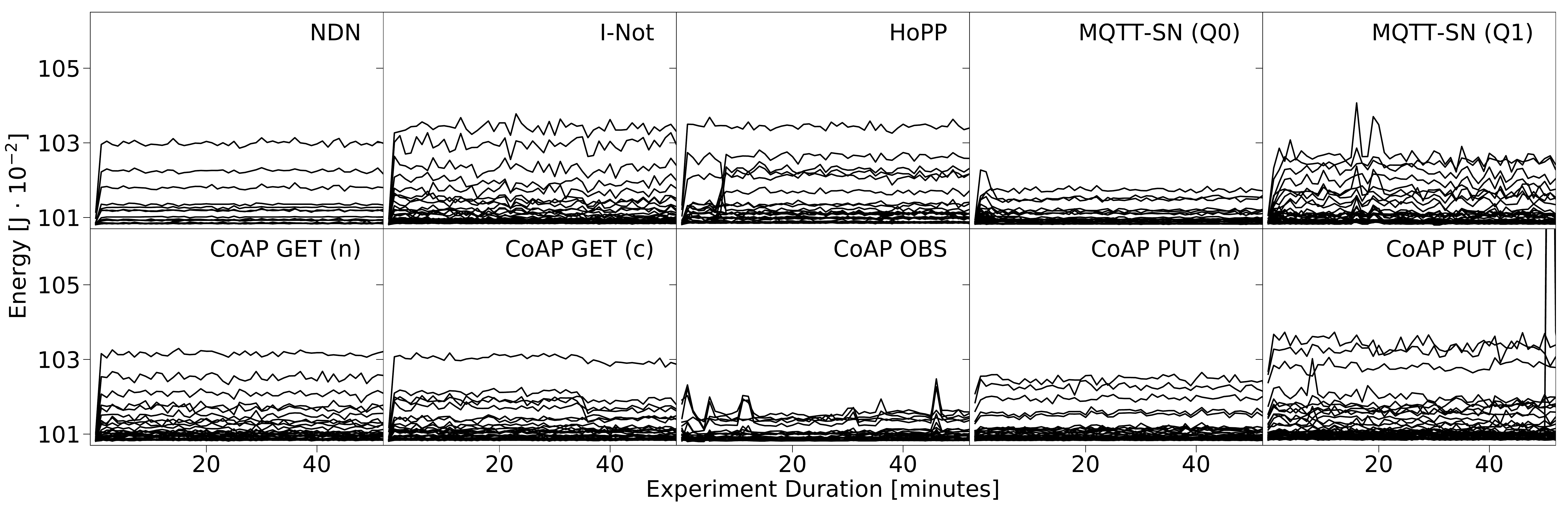}
     \caption{Energy consumption over time for each node in the topology using a 15 s publishing interval.}
      \label{fig:energy}
\end{figure*}

Time series of nodal energies are plotted in Figure \ref{fig:energy} for each protocol during the course of the experiment. Immediately we observe  the tree topology pattern in all graphs: The root node prominently consumes a multiple of  leaf node energies, and intermediate forwarders differentiate according to tree ranks in between. It is noteworthy that the  routing topology did not rearrange during the measurement period. A varying use of routing trees could gradually balance the uneven energy needs.

Aside from topological effects, distinct protocol signatures become visible. While all energy curves fluctuate due to temporal variations and local retransmissions, some protocols show significant amplitudes from local disorder and repair.
Reliable MQTT (Q1) exhibits a peak of recovery after an initial period of loss with depleted energy level on some branch, and a high number of pronounced peaks otherwise.

HoPP experiences a handover in energy load at about eight minutes. This follows its ability of dynamically switching to a more reliable uplink path without rebuilding the topology.
HoPP and NDN admit rather steady and smooth energy gradients, since they mainly rely on local repairs (or caching).
In contrast, I-Not as a protocol without in-network caching support requires more hop-wise retransmissions and must be considered energy-wise expensive.   

Unreliable protocols such as MQTT (Q0) and CoAP (n) repeatedly show valleys in energy curves, since packets lost early on the path relieve the  burden of forwarding to the remainders. Delivery failures in CoAP GET (c) as already known from Figure \ref{fig:link_stress} lead to some drops in energy, as well.
MQTT (Q0), CoAP OBS and CoAP PUT (n) consume the least energy, which is not surprising for these lean protocols without loss recovery.  

Viewing  link-stress (Fig.  \ref{fig:link_stress}) and energy flow (Fig. \ref{fig:energy}) conjointly, a rather clarifying view on the operational conditions of the protocols emerges. Some protocols remain lean and undemanding while delivering only a restricted service (e.g., CoAP OBS and PUT (n)), others are steady, predictable and run at full service (e.g., NDN and HoPP), and some protocols really struggle in this IoT-typic environment (e.g., MQTT (Q1), CoAP PUT (c), and I-Not).

\section{Related Work}
\label{sec:related_work}

\subsection{ICN and IoT}

The benefits of ICN/NDN in the IoT have been analyzed mainly from three angles.
\one design aspects~\cite{bmhsw-icnie-14,pf-britu-15,sylzz-nacce-15,mwt-tucin-16,arxpp-kip-17}, \two architecture work~\cite{g-ainai-17,szsmb-avdir-17}, and \three use cases~\cite{bgnt-sieoc-13,acim-icnis-15,srs-sndnt-15,gkslp-inii-17}.
To support experimental evaluation, several implementations have become available, including CCN-lite \cite{ccn-lite} on RIOT \cite{bhgws-rotoi-13,bghkl-rosos-18} and on Contiki \cite{alw-defsc-16}, and NDN on RIOT~\cite{saz-dinps-16}.
The objective of this paper is not to present an additional ICN implementation for the IoT but to reuse common stacks.
With this we contribute to more reliability of existing software as extensive usage helps to find bugs.

The evaluation of NDN protocol properties in the wild includes the exploitation of NDN communication patterns to improve wireless channel management~\cite{habsw-itpla-16,hbswa-litnc-17} as well as data delivery on the network layer~\cite{bmhsw-icnie-14}.
Comparison to common IoT network stacks at the transport layer (in particular UDP) is not available.
In this paper, we close the gap towards the application layer and analyze common application protocols (\ie MQTT and CoAP) compared to intrinsic network layer characteristic provided by NDN.

\subsection{Interoperation and Adoption of CoAP and MQTT in ICN}

Implementing CoAP on top of ICN has been proposed to enable full features of CoAP~\cite{filhp-coi-16,spp-cccn-17}, such as support of group communication and delay-tolerant communication~\cite{ilf-oiro-17}.
These concepts have been showcased in building management systems~\cite{fxpil-iecei-17}.
In contrast to the integration of CoAP into ICN, an MQTT-to-CCN gateway was proposed to allow for interoperation between CCN IoT devices and the current Internet~\cite{alw-defsc-16}.
A dedicated rendezvous point to discover resources and to bridge between IP-based MQTT subscribers and NDN sensors was introduced in~\cite{qfgac-itdii-17}.
Note that our work differs from those research as we assess the performance of CoAP, MQTT, and NDN in their original deployment scenarios, instead of focusing on interoperability use cases.
This helps to identify intrinsic protocol characteristics.

\subsection{Performance evaluation of CoAP and MQTT}

The performances of CoAP and MQTT have been studied from several perspectives over the last years~\cite{iou-aciii-17,dcjb-scpit-18}.
Very early work analyzed the interoperability of specific CoAP implementations \cite{lhk-iaitc-12,vpafb-coapl-12} without performance evaluation.
Later, CoAP implementations have been assessed in comparison to HTTP \cite{lmc-tncap-13} or on different hardware architectures~\cite{kh-bitd-14}.
MQTT was evaluated in \cite{esmcb-acpsm-15}.
Thangavel \emph{et al.}~\cite{tmvht-pemcc-14} proposed a common middleware to abstract from CoAP and MQTT. 
Based on this middleware, CoAP and MQTT were evaluated in a single-hop wired setup.
In emulation, MQTT and CoAP have been studied in the context of medical application scenario~\cite{ck-peipu-16}.
A holistic analysis of MQTT and CoAP in a consistent experimental setting including low-end IoT devices is missing.
In particular, no detailed comparison to NDN is provided.

\section{Conclusions and Outlook}
\label{sec:c+o}

This paper presented extensive experimental analyses to answer the question which of the common protocols MQTT, CoAP, or NDN is best suited for transferring IoT data from constrained devices. We found that for simple, single-hop topologies the protocol families examined in this paper  behave similar, but lean push protocols such as MQTT-SN and CoAP Observe operate fastest, at lowest energy consumption, and most network-friendly. 

In challenged multi-hop scenarios, though, the results quickly turn tides into a differentiated view between  protocols that operate in host-to-host semantic and those acting per link traversal. NDN and NDN-HoPP can both enfold their hop-wise  transfer features in balancing flows that reliably deliver data without the need for remarkable retransmission rates. This is in significant contrast compared to common UDP-based IoT application layer protocols that do not benefit from underlying flow control. 

While NDN is susceptible of overflowing PIT states in unscheduled publishing scenarios, NDN-HoPP handles such notification events without any performance flaw. In contrast, all IP-based protocols and also the NDN Interest Notification quickly struggle in challenging regimes, either by losing or by repeating packets at large scale. 

Our overall findings clearly indicate that lean and simple protocols such as MQTT and CoAP Observe can enfold its efficiencies in relaxed environments with low error rates. Challenged networks, though, will quickly degrade their performance to a minimum. In disruptive environments, protocol performance improves with operations confined to the local link: Hop-by-hop transfer with intermediate caching notably increases reliability and reduces corrective actions, which jointly grants efficient robustness. Dependable systems in challenged regimes should take advantage of corresponding solutions.

With these results, we hope to contribute insights to the community and to strengthen deployment in the constrained IoT. Our future work will concentrate on progressing, deploying, and measuring distributed data systems in the IoT domain that will grant operational insights from real-world deployment and at the same time foster an open, innovative, and robust Internet of Things.

\smallskip

\subsection*{A Note on Reproducibility}
We explicitly support reproducible research~\cite{acmrep,swgsc-terrc-17}.
Our experiments have been conducted in an open testbed.
The source code of our implementations (including scripts to setup the experiments, RIOT measurement apps etc.) will be available on Github at \url{https://github.com/5G-I3/ACM-ICN-2018}.

\subsection*{Acknowledgments}
We thank our shepherd Dave Oran and the anonymous ICN reviewers for their careful feedback and constructive guidance, which significantly helped to improve the paper.
This work was supported in parts by the German Federal Ministry of Research and Education within the project {\em I3: Information-centric Networking for the Industrial Internet}.

\balance

\bibliographystyle{ACM-Reference-Format}

\bibliography{own,rfcs,ids,theory,complexity,layer2,internet,transport,overlay,hypermedia,vcoip,visualization,security,ngi,manet,meta}
\end{document}